\documentclass[aps,prd,amsmath,amssymb,preprintnumbers,groupedaddress,nofootinbib,notitlepage]{revtex4-1}
\usepackage{tikz}
\usepackage{slashed}
\usepackage{hyperref}
\usepackage{verbatim}
\usepackage[normalem]{ulem}
\usepackage{wasysym}
\usepackage[utf8]{inputenc}

\newcommand{\cP}{\mathcal{P}}

\newcommand{\msun}{M_{\astrosun}}

\newcommand{\PR}{\mathcal{P}_\mathcal{R}}
\newcommand{\Pdelta}{\mathcal{P}_\delta}
\newcommand{\Ph}{\mathcal{P}_h}
\newcommand{\PPhi}{\mathcal{P}_\Phi}
\newcommand{\sRM}{\sigma_{R_M}}
\newcommand{\MPBH}{M_{\rm PBH}}
\newcommand{\eg}{{\it e.g.}}

\def\eg{{\it e.g.}}

\begin{document}
\setlength{\baselineskip}{0.22in}

\preprint{MCTP-16-33}
\title{Inflationary theory and pulsar timing investigations \\ of primordial black holes and gravitational waves}
\author{Nicholas Orlofsky$^a$, Aaron Pierce$^a$, James D. Wells$^{a,b}$\vspace{0.1in}} 
\vspace{0.2in}
\affiliation{$^a$Michigan Center for Theoretical Physics (MCTP)\\ Department of Physics, 
University of Michigan, Ann Arbor, MI 48109 USA \\ \\
$^b$Deutsches Elektronen-Synchrotron (DESY), Notkestra\ss e, D-22607, Hamburg, Germany \\ }

\vspace{5.9in}
\date{\today}

\begin{abstract}
The gravitational waves measured at LIGO are presumed here to come from merging primordial black holes. We ask how these primordial black holes could arise through inflationary models while not conflicting with current experiments. Among the approaches that work, we investigate the opportunity for corroboration through experimental probes of gravitational waves at pulsar timing arrays.  We provide examples of theories that are already ruled out, theories that will soon be probed, and theories that will not be tested in the foreseeable future.  The models that are most strongly constrained are those with a relatively broad primordial power spectrum.
\end{abstract}
\maketitle

\section{Introduction}
The LIGO detection of gravitational waves (GWs) resulting from the merger of black holes with masses $\sim 10$ to $30 \msun$ (where $\msun$ is a solar mass) \cite{Abbott:2016blz,Abbott:2016nmj} was an important validation of general relativity.  In addition, the merger rate observed by LIGO may be consistent with primordial black holes (PBHs) (for PBH reviews, see, \eg,  \cite{Carr:2009jm,Carr:2016drx}) making up all \cite{Bird:2016dcv} or a fraction $\sim 10^{-3}$ \cite{Sasaki:2016jop,Eroshenko:2016hmn} of the observed dark matter (DM) density. In this study, we investigate the consequences for theory and experiment by assuming the PBH merger explanation.

Primordial black holes within this mass range are subject to several constraints, especially if the fractional relic abundance $f=\Omega_{\rm PBH}/\Omega_{\rm DM} \gtrsim 0.1$.  Microlensing measurements \cite{Tisserand:2006zx,Wyrzykowski:2011tr,Allsman:2000kg} constrain black hole masses $\MPBH \lesssim 30 \msun$, with the strongest constraints coming at $\MPBH \lesssim \msun$.  Constraints from the ultra-faint dwarf Eridanus II \cite{Brandt:2016aco} give complementary bounds for $\MPBH \gtrsim 5$ to $30 \msun$, depending on assumptions about the density and velocity dispersion of DM.  These constraints when taken together may allow a monochromatic PBH mass spectrum at $\MPBH=30 \msun$ or allow an extended distribution with $f=\Omega_{\rm PBH}/\Omega_{\rm DM} \lesssim 0.1$ over a wide range of masses around this.  Bounds from these experiments on extended mass spectra are discussed in \cite{Carr:2016drx,Green:2016xgy}.  Bounds may also be placed on PBHs with mass $\MPBH \sim 1$ to $1000 \msun$ moving relative to pulsar lines of sight \cite{Schutz:2016khr}.  In addition to these, there are strong bounds from WMAP and FIRAS \cite{Ricotti:2007au} and Planck \cite{Chen:2016pud} that may limit $f \lesssim 10^{-2}$ to $10^{-4}$ around these PBH masses, though these bounds have been disputed \cite{Bird:2016dcv}.  Such a low abundance, nevertheless, may be consistent with the observed rate at LIGO if the higher merger rate estimates such as those given in Refs.~\cite{Sasaki:2016jop,Eroshenko:2016hmn} obtain.

Let us comment on estimations of PBH merger rates found in the literature.  In calculating the merger rate, the authors of Ref.~\cite{Bird:2016dcv} estimated binary formation inside DM halos today, while the authors of Refs.~\cite{Sasaki:2016jop,Eroshenko:2016hmn} presumed that binaries formed primordially.  In the primordial case, they estimated the probability for pairs of PBHs to gravitate enough to decouple from Hubble expansion (see also earlier work in Refs.~\cite{Nakamura:1997sm,Ioka:1998nz}).  The estimated rate for primordially produced binary mergers is much higher than the rate of mergers from binaries formed in halos, so the mergers of primordially produced binaries may be expected to dominate.  

However, there are several assumptions and quantities that can only be calculated numerically that lead to uncertainty in the rate of mergers from primordially formed binaries.  In isolation, a primordial PBH pair would merge essentially immediately after they decouple from the Hubble flow, with a time scale set by the gravitational free-fall time.  However, other PBHs may perturb the binary system by creating tidal forces that lead the pair to form eccentric orbits.  Once in this eccentric orbit, the time scale to merge via the emission of gravitational radiation is significantly prolonged. To estimate this new infall time, the semi-major and -minor axes for these binaries must be computed, a point on which the authors of Refs.~\cite{Ioka:1998nz,Eroshenko:2016hmn} disagree numerically (although in Ref.~\cite{Ioka:1998nz} a more realistic Poisson probability distribution was used compared to the flat distribution in \cite{Sasaki:2016jop,Eroshenko:2016hmn}).

Notably, no uncertainties are given for these calculations.  To estimate the theoretical uncertainties, we vary the parameters of the model within an order of magnitude.\footnote{Specifically, we vary $\alpha$ and $\beta$ as defined in \cite{Ioka:1998nz} between 0.1 and 10 and use a Poisson probability distribution.}
The uncertainty of the merger rate translates into a range of PBH DM fractions $f$ consistent with the LIGO rate from $f \sim 10^{-3}$ to $10^{-1}$.  
The theoretical uncertainties that may be included in this estimate include those discussed in Ref.~\cite{Ioka:1998nz}, such as the treatment of angular dependence, three-body collisions, other fluctuations beyond the three bodies considered, initial conditions, and radiation drag.  Additionally, all these calculations assume an idealized monochromatic mass spectrum of PBHs.  Finally, there is a further question of whether these binaries, which initially have major axes $\gtrsim$ the size of our solar system, would survive the process of halo formation.  These questions require further detailed numerical study beyond the scope of this work.  An interesting observation is that this collection of binary merger assumptions appears inconsistent with PBH DM fraction $f=1$.
For our purposes, we simply note that the present theoretical and experimental uncertainties may allow a wide range of PBH abundance.  For concreteness, we focus on the case where PBHs make up all of the DM, which might obtain if the above mechanism were ineffective (and in which case the mechanism of   Ref.~\cite{Bird:2016dcv} might dominate), but we will see that reducing the PBH abundance by even several orders of magnitude will have only a modest impact on the primary experimental probe we investigate in this paper and now introduce.

If these black holes were produced primordially, they would have resulted from the collapse of large density perturbations.  While these would be sourced as scalar perturbations, because of their size they can lead to nontrivial tensor perturbations at second order in cosmological perturbation theory \cite{Tomita:1967,Matarrese:1992rp,*Matarrese:1993zf,Matarrese:1997ay,Noh:2004bc,Carbone:2004iv}.  A key observation is that PBHs in the mass range detected by LIGO generate tensor perturbations that may be detected as gravitational waves at pulsar timing arrays (PTAs) \cite{Detweiler:1979wn}.  

To make a clear distinction from the GWs  detected at LIGO, we call the GWs probed at PTAs ``secondary gravitational waves" (SGWs). SGWs that are correlated with the LIGO GW signal may or may not be detectable at PTAs depending on the details of the formation mechanism of the PBHs.
For example, it has been observed that for scalar perturbations that are highly peaked---behaving essentially like Dirac $\delta$ functions---PTA probes of SGWs already exclude the formation of PBHs with masses in the range $10^{-2} \msun \lesssim \MPBH \lesssim 10 \msun$ \cite{Saito:2008jc,Saito:2009jt,Bugaev:2010bb}.  However, constraints have not been applied to explicit models for forming PBHs with nonidealized scalar power spectra until recently in Refs.~\cite{Garcia-Bellido:2016dkw,Inomata:2016rbd}.

In this paper, we discuss production mechanisms that may give rise to a somewhat narrow spectrum of $\sim 10$ to $30 \msun$ PBHs that make up some or all of the dark matter and are consistent with the LIGO GW signal.  We place constraints on these models using PTA sensitivities to SGWs.  Models with strongly peaked primordial spectra produce SGWs that can be probed in the near future by PTAs.  Meanwhile, models with extended primordial spectra are already excluded by present PTA data.  However, some models that explain PBH formation consistent with LIGO will not be probed by present or future PTA experiments.  The most important limiting factor for probing these models is PTA observing time.
SGW detection by a PTA would not only bolster the case for the merging black holes detected by LIGO as being formed primordially, but also provide insight into the physics of the very early universe.  

The remainder of this paper is organized as follows.  In Sec.~\ref{sec:formalism} we review the calculation of PBH and SGW spectra from a primordial scalar spectrum.  Section \ref{sec:delta} gives a demonstration of the SGW spectrum for an idealized $\delta$ function primordial spectrum.  Section \ref{sec:models} discusses explicit modes for PBH formation and places bounds on their resulting SGW spectra.  In Sec.~\ref{sec:discussion} we discussion assumptions and uncertainties in our calculations and their effects on the bounds.  We conclude in Sec.~\ref{sec:conclusion}.

%%%%%%%%%%%%%%%%%%%%%
\section{PBH and GW spectra from a primordial scalar spectrum}
\label{sec:formalism}

We consider a primordial curvature perturbation spectrum $\PR(k)$ whose form is determined by early universe dynamics.  Our approach is to ensure that PBHs are formed in the right mass range and with the right abundance to explain the LIGO GW signal. The spectrum then has the potential to give rise to a background of stochastic SGWs, whose rate and strength depend on details of the inflationary theory.  These details will be described in Sec.~\ref{sec:models}. Here we review the formalism for determining PBH formation and SGWs from a generic perturbation spectrum. 

First, let us consider PBH formation.  The curvature perturbations result in density perturbations, which during radiation domination are described by power spectra\footnote{More properly, $\Pdelta(k)=\frac{16}{3} \left( \frac{k}{a H} \right)^2 j_1^2\left(\frac{k}{\sqrt{3}a H}\right) \PR(k)$ \cite{Josan:2009qn}, where $j_1$ is a spherical Bessel function, though the difference mainly appears on sub-horizon scales that are suppressed in (\ref{eqn:sigmaRM}) by the window function $W(k R)$.} 
$\Pdelta(k)= \frac{4}{9} \left( \frac{k}{a H} \right)^4 \PPhi(k) = \left( \frac{4}{9} \right)^2 \left( \frac{k}{a H} \right)^4 \PR(k)$ for the matter perturbations $\delta=\delta\rho/\rho$, Bardeen potential $\Phi$, and curvature $\mathcal{R}$.  Here, $a$ is the scale factor and $H=\frac{1}{a}\frac{da}{dt}$ is the Hubble parameter.  When the perturbations are large enough, an overdense region can collapse into a PBH when the overdensity reenters the horizon at scale $k_f$, resulting in a PBH approximated to have a horizon mass
\begin{equation}
\label{eqn:MPBH}
M_{\rm PBH}=\frac{4 \pi}{3} \rho_r H^{-3} \simeq 10 \msun \left(\frac{g_*}{100}\right)^{-1/6}\left(\frac{\text{pc}^{-1}}{k_f}\right)^{2},
\end{equation}
where the universe is assumed to be radiation dominated with energy density $\rho_r$, and $g_*$ is the number of relativistic degrees of freedom.

Assuming these perturbations are Gaussian, the energy fraction of PBHs with mass in the interval $(M,M+dM)$ at their formation time $t_f$ can be calculated using Press-Schechter formalism \cite{Press:1973iz,Kawasaki:2016pql},
\begin{equation} \label{eqn:beta}
\beta(M)= \frac{d}{d\log M} \frac{\rho_{\rm PBH}(t_f)}{\rho_{\rm tot}(t_f)} 
= 2 \int_{\delta_c}^{\infty} d\delta \frac{1}{\sqrt{2\pi} \sigma} e^{-\frac{\delta^2}{2 \sigma^2}} \\
= {\rm Erfc}\left(\frac{\delta_c}{\sqrt{2} \sRM}\right),
\end{equation}
where \cite{Young:2014ana}
\begin{equation} \label{eqn:sigmaRM}
\sRM^2=\int \frac{dk}{k} W(k R_M)^2 \Pdelta(k)
=\int \frac{dk}{k} W(k R_M)^2 \left(\frac{4}{9}\right) (k R_M)^4 \PPhi(k)
\end{equation}
is the variance for the Gaussian probability distribution for primordial density perturbations on the length scale $R_M$.  Here, $W(x)=e^{-x^2/2}$ is a Gaussian window function.  
We take the threshold for PBH formation $\delta_c=1/3$ \cite{Carr:1975qj} in Eq.~(\ref{eqn:beta}) which will result in a conservative estimate of the SGW abundance, though higher values of, \eg,~$\delta_c \simeq 0.45$ have also been indicated \cite{Musco:2004ak,*Musco:2008hv,*Musco:2012au}, and nonsphericity effects \cite{Musco:2012au,Sheth:1999su,Kuhnel:2016exn,Bond:1993we} can make $\delta_c$ higher still.  See the Discussion section for the effect of these different choices.  Finally, the relic abundance of PBH today is given by
\begin{equation}
\frac{d}{d\log M} \Omega_{\rm PBH}h^2 \simeq 2 \times 10^7 \beta(M) \left(\frac{g_{*,i}}{106.75}\right)^{-1/4} \left(\frac{M}{\msun}\right)^{-1/2}.
\end{equation}

Next, let us consider SGW production.  By definition here, SGWs are GWs that are produced at second order in perturbation theory and could be probed by PTAs.  Precisely because the scalar power spectrum must be large to produce PBHs, these secondary tensor modes may be detectable.

The second-order tensor power spectrum can be calculated from the scalar power spectrum using \cite{Ananda:2006af,Baumann:2007zm,Bugaev:2009zh,Bugaev:2010bb,Saito:2009jt,Alabidi:2012ex},\footnote{See Refs.~\cite{Ananda:2006af,Alabidi:2012ex} for variable changes to make this computationally simpler, where we have resolved some inconsistencies and ambiguities.}
\begin{equation} \label{eqn:Ph}
\Ph(k,\eta)=\int_{0}^{\infty}d\tilde{k} \int_{-1}^{1}d\mu~\PPhi(|\mathbf{k}-\mathbf{\tilde{k}}|)\PPhi(\tilde{k})\mathcal{F}(k,\tilde{k},\mu,\eta),
\end{equation}
where
\begin{align}
\label{eqn:PhF}
\mathcal{F} (k,\tilde{k},\mu,\eta)= 
\frac{(1-\mu^2)^2}{a^2(\eta)}\frac{k^3\tilde{k}^3}{|\mathbf{k}-\mathbf{\tilde{k}}|^3} \int_{0}^{\eta} d\eta_1 ~ a(\eta_1) g_k(\eta,\eta_1) f(\mathbf{k},\mathbf{\tilde{k}},\eta_1) 
\int_{0}^{\eta} d\eta_2 ~ a(\eta_2) g_k(\eta,\eta_2) \left[f(\mathbf{k},\mathbf{\tilde{k}},\eta_2)+f(\mathbf{k},\mathbf{k}-\mathbf{\tilde{k}},\eta_2)\right],
\end{align}
with
\begin{align}
f(\mathbf{k},\mathbf{\tilde{k}},\eta)= 12 \Phi(\tilde{k}\eta) \Phi(|\mathbf{k}-\mathbf{\tilde{k}}|\eta)+ 8\eta\Phi(\tilde{k}\eta) \Phi'(|\mathbf{k}-\mathbf{\tilde{k}}|\eta)
+ 4\eta^2 \Phi'(\tilde{k}\eta) \Phi'(|\mathbf{k}-\mathbf{\tilde{k}}|\eta).
\end{align}
Here, $\mu=\mathbf{k} \cdot \mathbf{\tilde{k}}/(k \tilde{k})$ and $\eta$ is the conformal time.
The Bardeen potential $\Phi$ during radiation domination (RD) (after dropping the decaying mode) is
\begin{equation}
\Phi(\mathbf{k},\eta)=\frac{A(\mathbf{k})}{(\sqrt{w} k \eta)^2} \left(\frac{\sin(\sqrt{w}k\eta)}{\sqrt{w}k\eta}-\cos(\sqrt{w}k\eta)\right),
\end{equation}
with the equation of state $w=1/3$ during RD.  Its power spectrum is defined by
\begin{equation}
\left<\Phi(\mathbf{k})\Phi(\mathbf{k}')\right>=\frac{2\pi^2}{k^3} \delta^3(\mathbf{k}+\mathbf{k}') \PPhi(k).
\end{equation}
The Green's function in Eq.~(\ref{eqn:PhF}) is
\begin{equation}
g_k(\eta,\eta')=\frac{\sin(k(\eta-\eta'))}{k}.
\end{equation}
Finally, the relic abundance of gravitational waves can be calculated in terms of Eq.~(\ref{eqn:Ph}) as \cite{Maggiore:1999vm,*Maggiore:1900zz,Bugaev:2009zh}\footnote{References \cite{Ananda:2006af,Baumann:2007zm,Saito:2009jt,Alabidi:2012ex} give a range of differing values for this expression with which we do not agree.}
\begin{equation}
\frac{d}{d \ln k}\Omega_{\rm SGW}(k,\eta) = \frac{1}{12}\left(\frac{k}{a(\eta)H(\eta)}\right)^2 \Ph(k,\eta).
\end{equation}
Since $\Omega_{\rm SGW}$ scales as radiation, it is convenient to evaluate this quantity at matter-radiation equality (denoted by the subscript ``eq'') and then scale to today (denoted by the subscript ``0''), giving approximately
\begin{equation}
\frac{d}{d \ln k}\Omega_{\rm SGW}(k,\eta_0) \simeq \frac{1}{12}\frac{1}{1+z_{\rm eq}} (k \eta_{\rm eq})^2 \Ph(k,\eta_{\rm eq}),
\label{eq:omegasgw}
\end{equation}
where $z$ is the redshift.

Finally, let us review how experimental searches connect with gravitational wave abundance. Gravitational wave experiments typically quote results in terms of the characteristic strain $h_c$, which is related to an abundance of stochastic gravitational waves by \cite{Maggiore:1999vm,*Maggiore:1900zz}
\begin{equation}
\frac{d}{d \ln k}\Omega_{\rm GW}(k=2\pi f,\eta_0)= \frac{2\pi}{3 H_0^2}f^2 h_c^2(f).
\end{equation}
The $\Omega_{\rm SGW}$ abundance computed in Eq.~(\ref{eq:omegasgw}) can be directly translated to characteristic strain constraints using this formula.

%%%%%%%%%%%%%%%%%%%%%%%%%%%%%%%%%
\section{Idealized delta function spectrum}
\label{sec:delta}

Perhaps the simplest model for PBH production is to assume a sharp narrow spike in $k$ space in the scalar power spectrum.  For a narrow enough spike, this can be approximated by a $\delta$ function \cite{Saito:2009jt}.  While a $\delta$ function is not physical, it is useful to consider as a mathematical construct. 

\begin{figure}
\includegraphics[width=0.5\columnwidth]{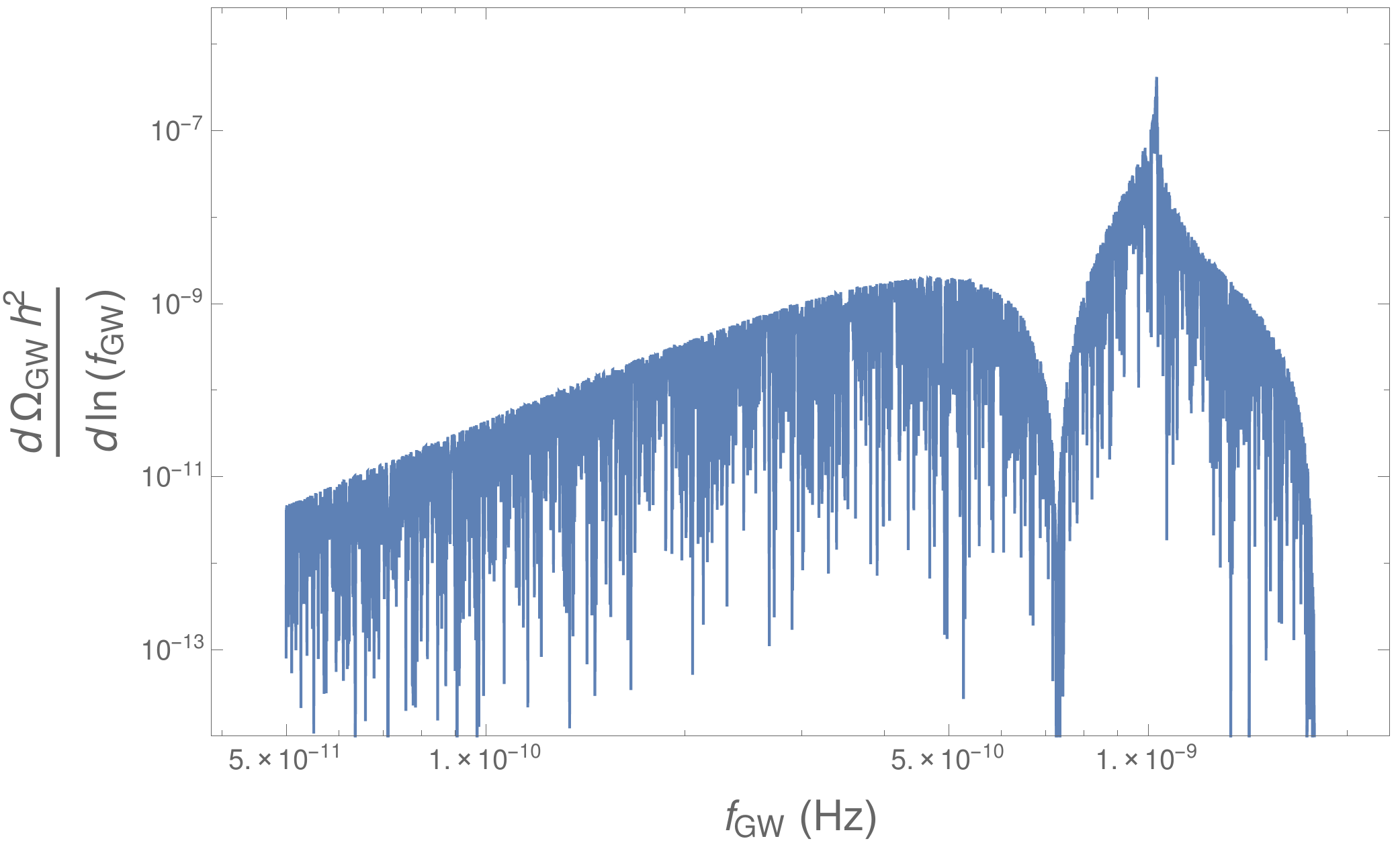}
\caption{Gravitational wave abundance for an idealized delta function scalar spectrum peaked at the scale $k_f$ corresponding to PBH mass $30\msun$ according to Eq.~(\ref{eqn:MPBH}) and normalized so that $\Omega_{\rm PBH}=\Omega_{\rm DM}$}
\label{fig:GWdelta}
\end{figure}

The SGW spectrum for a $\delta$-function scalar spectrum is plotted in Fig.~\ref{fig:GWdelta}.  For reference, the $\delta$ function is chosen to be peaked at a scale $k_f$ corresponding to a horizon PBH mass of $30\msun$ (see Eq.~(\ref{eqn:MPBH})), and its amplitude is chosen so that PBHs that form from this spectrum make up all of the observed DM abundance.   

A constructive interference between $g_k$ and $f(\mathbf{k},\mathbf{\tilde{k}},\eta)$ in Eq.~(\ref{eqn:PhF}) leads to a resonance at $k=2 \sqrt{w}k_f$ where the amplitude continues to grow at late times, and there is a zero at $k=\sqrt{2w}k_f$.  The spectrum extends up to $k=2 k_f$ where the incoming scalar modes are aligned.

Of course, physical spectra will be extended.  Let us briefly discuss the changes to the PBH and SGW spectra as we go to more extended primordial scalar spectra.
Regarding the PBH spectrum, note that PBH formation is exponentially sensitive to $\sRM$ and the integration in Eq.~(\ref{eqn:sigmaRM}) samples a somewhat narrow window in $k$ of $\PPhi(k)$.  Thus, PBHs predominately form near where $\PPhi(k)$ peaks.  Nevertheless, the integration in Eq.~(\ref{eqn:sigmaRM}) may lead the PBH mass spectrum to peak at a smaller or larger scale than $\PPhi(k)$ depending on the detailed shape of $\PPhi(k)$.  Critical collapse effects \cite{Gundlach:1999cu,*Gundlach:2002sx,Niemeyer:1997mt,*Niemeyer:1999ak,Kuhnel:2015vtw,Musco:2004ak,*Musco:2008hv,*Musco:2012au} will lead to further corrections to the peak mass---see comments in the Discussion section.
In regards to SGWs, the SGW abundance spectra will be smoothed out by the integral in Eq.~(\ref{eqn:Ph}) for extended $\PPhi(k)$, so features like the resonance and destructive interference in the $\delta$-function spectrum will not be present for extended spectra.  Additionally, the SGW abundance is only quadratically sensitive to $\PPhi(k)$ and depends on integration over a larger range of $k$ in Eq.~(\ref{eqn:Ph}).  This leads to an enhanced SGW abundance over a larger range in frequency for extended scalar spectra relative to a narrower spectrum.

%%%%%%%%%%%%%%%%%%%%%%%%%%%%%%%%%
\section{Models of primordial scalar spectra}
\label{sec:models}

\begin{figure*}
\mbox{\includegraphics[width=0.49\columnwidth]{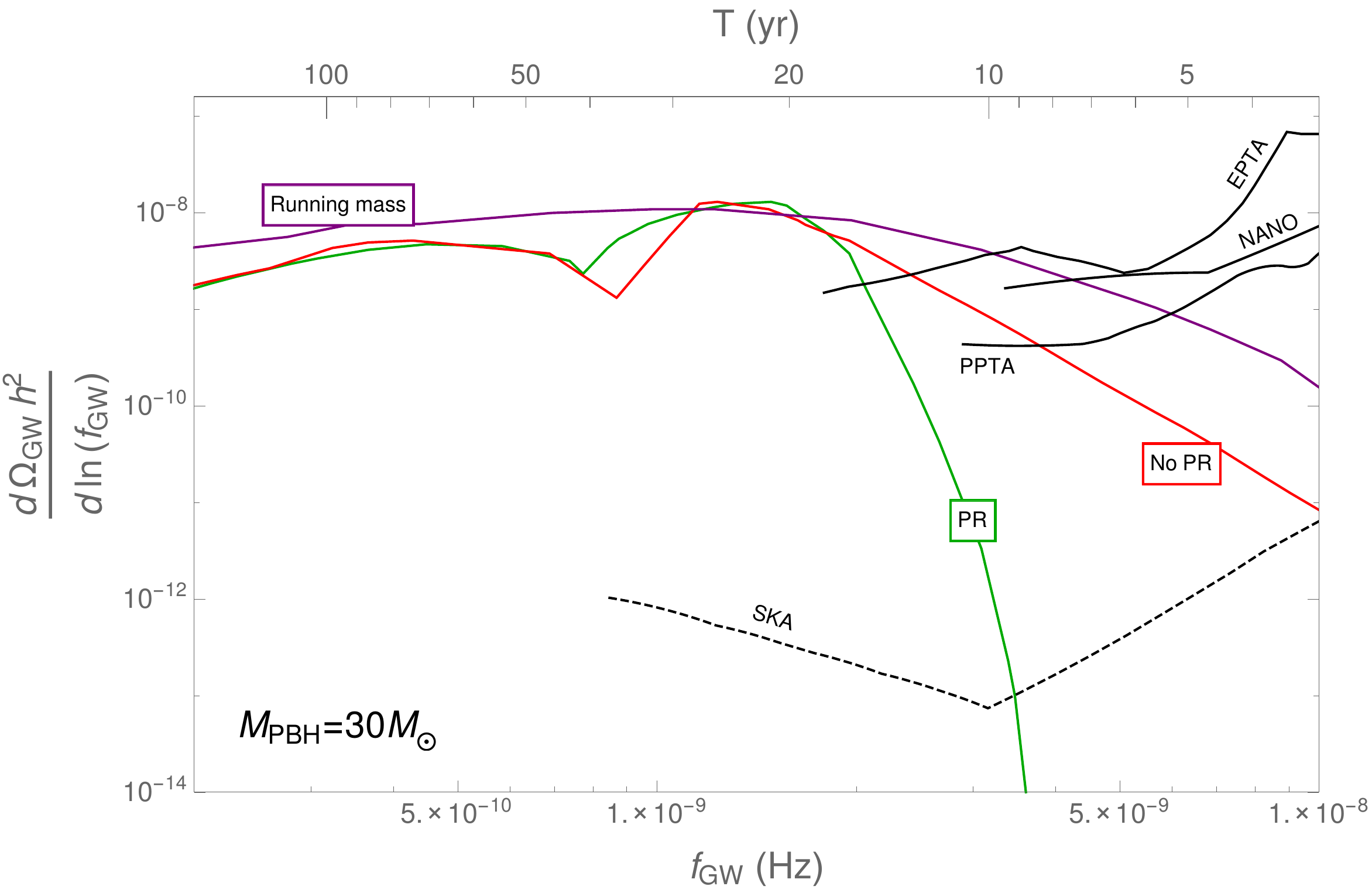}}
\mbox{\includegraphics[width=0.49\columnwidth]{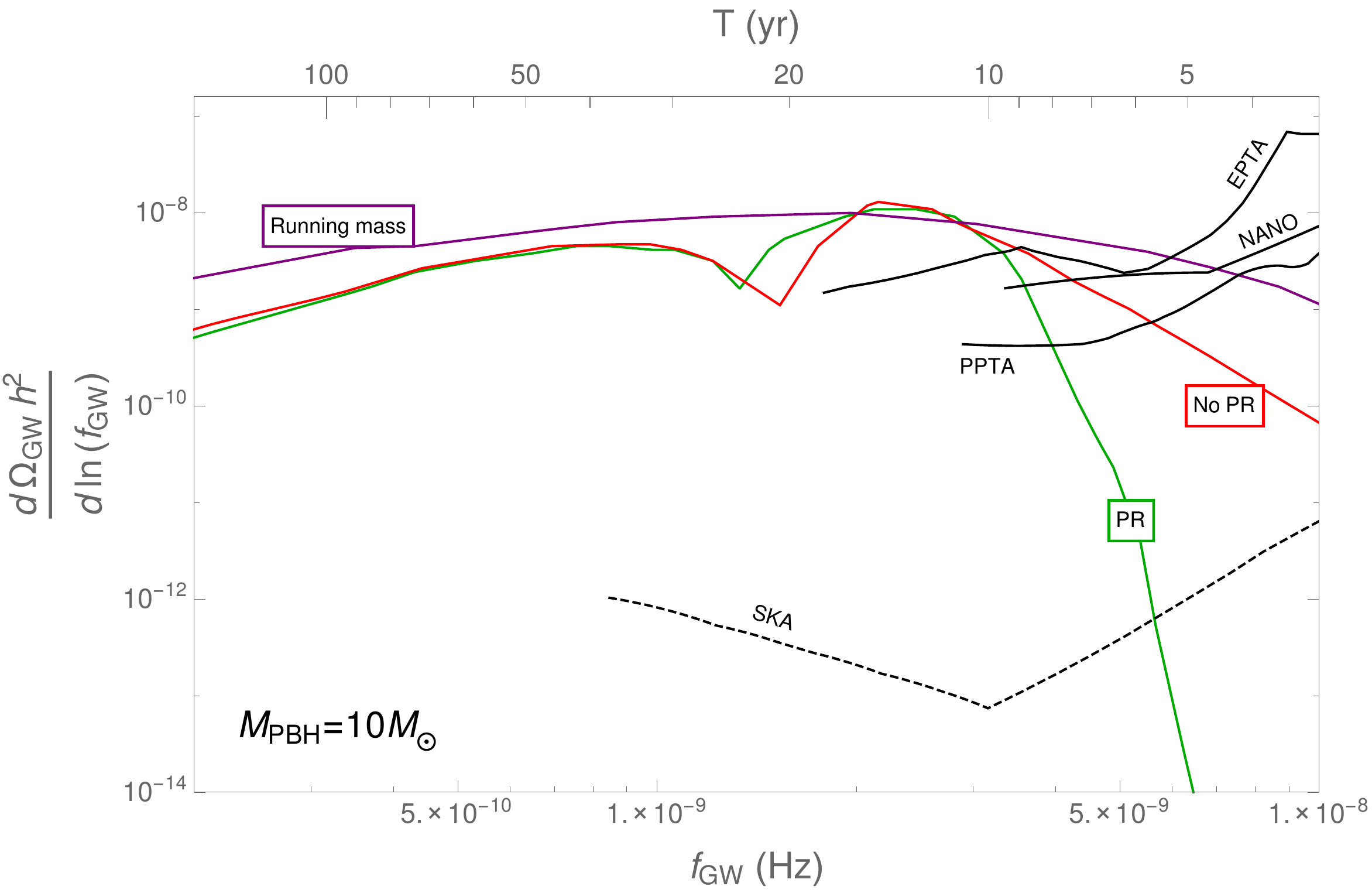}}
\caption{Gravitational wave abundance (envelopes) as a function of frequency assuming $\Omega_{\rm PBH}=\Omega_{\rm DM}$.  The PBH abundance spectrum is peaked at 30 (left) or 10 (right) $\msun$.  We display SGW for a top-hat spectrum with width set by expectations from parametric resonance (green ``PR'' curve), a red-tilted scalar spectrum with spectral index $n_s=-1$ supplemented by a cutoff at a minimum frequency (red ``No PR'' curve), and the spectrum from the running mass model (purple curve).  Black solid lines are current spectrum-independent bounds from EPTA (upper) \cite{Lentati:2015qwp}, NANOGrav (middle) \cite{Arzoumanian:2015liz}, and PPTA (lower) \cite{Shannon:2015ect}.  The black dashed line is a projection for bounds from SKA \cite{Janssen:2014dka}.  
The top axis indicates the approximate observing time $T$ to be sensitive to a given minimum frequency $f_{\rm min} \sim 1/T$.
}
\label{fig:GWspec}
\end{figure*}

There are many models of inflationary dynamics that can induce primordial power spectra giving rise to PBH formation.  Here, we review several classes of models capable of producing PBHs with large enough abundance at masses relevant to LIGO.  We do not attempt a full accounting of all models that have been proposed for PBH production.  Rather, we survey several models that predict different primordial spectra, allowing us to draw some general conclusions.  Other models of PBH production not considered here can be found in Refs.~\cite{Dolgov:1992pu,GarciaBellido:1996qt,Clesse:2015wea,Cheng:2016qzb,Garcia-Bellido:2016dkw,Blinnikov:2016bxu}.
For each model, we calculate the resulting SGW spectrum.  Results and present bounds from the European Pulsar Timing Array (EPTA) \cite{Lentati:2015qwp}, Parkes Pulsar Timing Array (PPTA) \cite{Shannon:2015ect}, and NANOGrav \cite{Arzoumanian:2015liz} experiments and projections for SKA \cite{Janssen:2014dka} are displayed in conjunction with these spectra in Fig.~\ref{fig:GWspec}.  The plots show envelopes of SGW spectra that correspond to a peak in the PBH mass spectrum at masses $M_{\rm PBH}=30~\text{and}~10 \msun$.  We show results for $\Omega_{\rm PBH}=\Omega_{\rm DM}$, but the SGW amplitude is only log-dependent on the PBH relic abundance.  Thus, results are only slightly changed for other choices.  For example, for a spectra peaking at $M_{\rm PBH}=30 \msun$, reducing the PBH abundance from $f=1$ (all the DM)  to $f=10^{-4}$ reduces the SGW relic abundance by a factor of $0.4$, which would have a minor effect on SGW detectability, whereas the LIGO rate may be accommodated in either case depending on the details of the binary formation and merger rate calculation as discussed in the Introduction.

%%%%%%%%%%%%
\subsection{Double inflation with parametric resonance}

A close approximation to a highly peaked $\delta$-function scalar spectrum can be realized with a period of parametric resonance after the end of inflation.  Oscillations of the inflaton can lead to specific modes being exponentially enhanced \cite{Kofman:1994rk}.  Since parametric resonance occurs after inflation has ended, the resonantly excited modes would not have a large enough length scale to produce PBHs at masses relevant to LIGO.  A solution is to have two periods of inflation, with parametric resonance occurring between the two periods.  The second inflation stretches the resonantly amplified modes to the relevant length scale.  A model of this type was given in Refs.~\cite{Kawasaki:2006zv,Kawaguchi:2007fz,Frampton:2010sw}, where a period of hybrid inflation \cite{Copeland:1994vg,Linde:1997sj} followed by new inflation \cite{Izawa:1996dv,Izawa:1997df} was constructed.  Such a model can be engineered to give a peak at any scale (by varying the length of the second inflation) and with any amplitude (depending on the relationship between the efficiency of the resonance and the decay width of the inflaton causing the resonance).

Because parametric resonance produces a sharply peaked scalar perturbation spectrum, it is the model that most closely mimics the $\delta$-function spectrum of the previous subsection.  However, even in this case, the $\delta$-function approximation is not quite applicable; parametric resonance at a scale $k$ results in a resonant band of width $\sim k$ \cite{Kofman:1997yn}.  Such a spectrum can be approximated more closely by a top hat \cite{Saito:2009jt}, and the resulting SGW spectrum is shown in Fig.~\ref{fig:GWspec} by a green curve.\footnote{In the notation of Ref.~\cite{Saito:2009jt}, the resonant band has amplitude $\PPhi \simeq A^2/(2\Delta)$ on the domain $|\ln (k/k_p)|<\Delta$.  For parametric resonance peaked at $k_p$ with width $k_p$, this implies $\Delta \simeq \sinh^{-1}(1/2) \simeq 0.48$.  PBHs are predominantly produced at the smallest masses within this window corresponding to the scale $k=k_p e^\Delta$.}
This more realistic spectrum does not have as pronounced a peak as the $\delta$-function case.  Furthermore, for a fixed peak mass of PBH production, the scalar power spectrum peak must be shifted to larger $k$ due to the integration in Eq.~(\ref{eqn:sigmaRM}), and therefore the SGW frequencies in this more realistic spectrum are larger than those of the $\delta$-function spectrum. 

At present, pulsar timing constraints have sensitivity to this model up to PBH mass spectra peaked at $\MPBH \lesssim 30 \msun$.  However, it should be noted that critical collapse effects will somewhat reduce the mass reach shown here; see details in the Discussion section.  Thus, this model cannot be definitively excluded at present as an explanation for the black holes observed by LIGO.  Note that although EPTA is not as sensitive to small GW abundance as the other experiments, its longer data collection time allows it to probe smaller frequencies which are critical here to detecting SGW signals.  Thus, even absent any gain in sensitivity to $h_c$, all that would be needed to probe larger masses is increased data collection time $T$.  The minimum frequency that can be probed is $f_{\rm min} \sim 1/T$; therefore, as observing time increases, the maximal mass probed will go as $M \propto T^2$ (see Eq.~(\ref{eqn:MPBH})).  We translate frequency to observing time along the top axis of Fig.~\ref{fig:GWspec}.

{\it Summary:} Double inflation with parametric resonance has enough flexibility to allow a narrow range of PBHs needed to produce the LIGO GW signal. SGW spectra for masses $\MPBH \lesssim 30 \msun$ are constrained before accounting for critical collapse effects.  An increase in PTA data collection time will enable detection of the SGWs for spectra peaked at masses greater than this, potentially detecting or excluding this mechanism as an explanation for LIGO GWs. 

%%%%%%%%%%%%%%%%%%

\subsection{Double inflation without parametric resonance}

Double inflation can produce a somewhat peaked spectrum even without parametric resonance.  The earlier period of inflation gives rise to perturbations observed in the cosmic microwave background (CMB) and large-scale structure.   This early inflation is thus bounded to have small perturbations $\delta \rho/\rho \simeq 10^{-5}$.  However, a later period of inflation can have much larger perturbations.  This can be constructed with a model similar to that of the previous subsection; hybrid inflation followed by new inflation \cite{Kawasaki:1997ju,Kawasaki:1998vx}.  The detectability of this model using PTAs was recently discussed in Ref.~\cite{Inomata:2016rbd}, with which our conclusions are in agreement.  To produce a large peak in the power spectrum, the model is designed so that the scale $k_*$ at which the second inflation starts roughly coincides with the scale relevant to PBH formation.
The power spectrum of the second inflation starts out very large but has a red spectral tilt ($n_s<1$).  Thus, the spectrum is $\cP_\mathcal{R}=\cP_\mathcal{R}(k_*) (k/k_*)^{n_s-1}$ for $k>k_*$ and $\cP_\mathcal{R} \ll \cP_\mathcal{R}(k_*)$ for $k<k_*$.  In this model, the slow-roll parameters satisfy $|\eta| \gg |\epsilon|$, so using the relationship $n_s-1=2\eta-6\epsilon$ and the slow-roll condition $|\eta|<1$ implies $-1 \lesssim n_s \lesssim 3$ \cite{Izawa:1996dv}.

For this model, given a peak PBH mass $M_{\rm peak}$, $k_*$ is smaller than the formation scale $k_f(M_{\rm peak})$ given by Eq.~(\ref{eqn:MPBH}).  This results from the integration in Eq.~(\ref{eqn:sigmaRM}), which is maximized when $R_M^{-1}>k_*$.  Thus, the SGW spectrum will be peaked at smaller frequencies in this model compared to the case of a $\delta$ function giving the same peak PBH mass.  This effect is more pronounced for larger $n_s$ corresponding to wider spectra, whereas for the minimum $n_s=-1$, $k_* \simeq k_f(M_{\rm peak})$.  On the other hand, the SGW abundance away from the peak scale $k_*$ is enhanced relative to narrower spectra.

Figure \ref{fig:GWspec} shows the resulting SGW spectra as red curves assuming $n_s=-1$, the minimum consistent with slow roll. This choice gives the narrowest possible peak in the perturbation spectrum.  Larger values of $n_s$ would be more strongly constrained because they represent a more extended perturbation spectrum with larger resulting SGW abundance at high frequencies, which more than makes up for the necessary reduction in $k_*$ to ensure the peak PBH mass $M_{\rm peak}$ remains constant as $n_s$ increases.
Already, the EPTA and PPTA experiments exclude this model of PBH formation up to masses peaked at greater than $30 \msun$ due to the contributions to SGWs at high frequency.\footnote{Data from multiple frequencies can be combined to yield stronger bounds on extended spectra.  NANOGrav gives such a bound, but it is no stronger than the bin-by-bin exclusion for the sharply falling spectrum considered here.}
However, it is possible that critical collapse effects may still narrowly allow this model.

{\it Summary:} This double inflation model can produce PBHs in a narrow mass range with an abundance consistent with the rate of binary mergers observed by LIGO.  However, because of its more extended spectrum, this model produces SGWs across a wider range of frequencies.  Thus, present PTA experiments are more sensitive to this model than the previous and exclude this as an explanation for LIGO GWs up to the details of critical collapse effects.

%%%%%%%%%%%%%%%%%%%%%%%%%%%%%
\subsection{Running-mass model}

The running-mass model \cite{Stewart:1996ey,*Stewart:1997wg,Leach:2000ea,Drees:2011hb,*Drees:2011yz} supposes just one period of inflation, but with significant running of the spectral index.  This approach can achieve a large perturbation amplitude at scales much smaller than those relevant to the CMB (here denoted $k_0=0.05~{\rm Mpc}^{-1}$).  The amplitude at any point can be parametrized by,
\begin{equation}
\PR(k)=\PR(k_0) \left(\frac{k}{k_0}\right)^{n(k)-1},
\end{equation}
where,
\begin{equation}
n(k)= n_s(k_0)+\frac{1}{2!} \alpha_s(k_0) \ln \left(\frac{k}{k_0}\right) + \frac{1}{3!} \beta_s(k_0) \ln^2 \left(\frac{k}{k_0}\right)
+ \frac{1}{4!} \gamma_s(k_0) \ln^3 \left(\frac{k}{k_0}\right)+...
\end{equation}
Here, $\alpha_s$, $\beta_s$, and $\gamma_s$ are the runnings of $n_s$, giving sequentially higher derivatives of $n_s$ with respect to $\log k$.  When they are all allowed to vary, the parameters are bounded at $k=k_0$ as $\PR=(2.142 \pm .049) \times 10^{-9}$ \cite{Ade:2015xua}, $n_s = 0.9586 \pm 0.0056$, $\alpha_s=0.009 \pm 0.010$, and $\beta_s = 0.025 \pm 0.013$ \cite{Ade:2015lrj}.  To obtain a peaked spectrum, we will generally want one or both of $\alpha_s$ and $\beta_s$ to be positive, while $\gamma_s$ (or a higher-order term) must be negative.  The coefficients are chosen so that the peaking occurs for scales relevant for PBH formation, well outside of the region probed by the CMB.

For an explicit model of this type, see the above references.  We will simply set at $k=k_0$ the values $\PR=2.142 \times 10^{-9}$, $n_s=0.96$, and $\alpha_s = 0.009$ near their preferred experimental values.  We then vary $\beta_s$ and $\gamma_s$ to achieve a PBH fractional abundance vs.~mass spectrum that peaks at a given mass (here, as before, either 30 or $10 \msun$) and gives the desired total relic abundance equal to the DM abundance.\footnote{Specifically, $\beta_s=0.0903091~(0.0813319)$ and $\gamma_s=-0.0166717~(-0.0145293)$ for peak $\MPBH=30\msun~(10\msun)$.} 
Of note, this requires a tuning of several significant digits in the running parameters to obtain the correct peak position and amplitude in the primordial spectrum.  Additionally, for the masses of interest, $\beta_s$ is several standard deviations away from its measured value.

Another shortcoming is that generically $|\beta_s|$ should be suppressed by a factor of $(n_s-1)$ in relation to $|\alpha_s|$ if the third derivative of the inflaton potential is not much larger than the lower derivatives.  This is the case, \eg, for the explicit model in \cite{Drees:2011hb}.  However, to produce a large enough peak in the curvature spectrum for the PBH masses of interest, $|\beta_s| \gtrsim |\alpha_s|$ is required.   Constructing an explicit model to circumvent this generic suppression poses a challenge.

As in the previous subsection, PBH production occurs predominantly on scales where $\PPhi$ is largest, while SGWs are produced at all frequencies.  The result is plotted in purple in Fig.~\ref{fig:GWspec}.  While in principle stronger bounds could be placed by combining PTA data over many frequencies, even the sensitivity of the bin-by-bin exclusions surpasses the SGW abundance for this model.  Thus, present PTA experiments exclude this model for PBH production as an explanation for the LIGO events.

{\it Summary:} Like the double inflation without parametric resonance model of the previous subsection, the running mass model has an extended $\PPhi$ spectrum.  It can produce a somewhat narrow range of PBH masses in the LIGO window, though there are theoretical challenges with constructing such a model related to the size of the runnings.  The necessary runnings may also be inconsistent with present bounds from the CMB.  Even if these issues are ignored, this model leads to a large enough SGW abundance over a wide range of frequencies to exclude this model as an explanation for LIGO GWs.

%%%%%%%%%%%%%%%%%%%%%%%%%%%%%%%%
\subsection{Axion-curvaton model}

Unlike the previous models discussed, the axion-curvaton model \cite{Kasuya:2009up,Kawasaki:2012wr} (see also \cite{Kohri:2012yw})  supposes that primordial fluctuations on small scales are sourced after inflation from a separate curvaton field \cite{Lyth:2001nq,Lyth:2002my} that need not induce a second period of inflation.  The model consists of a complex field $\Phi=(\varphi/\sqrt{2}) e^{i \sigma/f_\sigma}$, where $\sigma$ is the curvaton.  Once $\varphi$ reaches the minimum of its potential at $\varphi=f_\sigma$ and begins oscillating about it, $\sigma$ becomes well defined.  After this point, corresponding to comoving scale $k_*$, the curvaton can induce a blue scalar spectrum ($n_s>1$) so that $\cP_\mathcal{R}=\cP_\mathcal{R}(k_*)(k/k_*)^{n_s-1}$ for $k<k_*$ and $\cP \ll \cP_\mathcal{R}(k_*)$ for $k>k_*$.  Here, a blue spectrum with $n_s \sim$ 2 to 4 can be obtained.  The power spectrum from the curvaton is constrained on large scales to be less than that observed to be coming from inflation---$\cP_\mathcal{R}(k) \lesssim 2 \times 10^{-9}$ for $k \lesssim$ Mpc$^{-1}$.

Because in this model the universe is assumed to always be radiation dominated, the curvaton, which redshifts as matter, decays before it dominates the universe's energy density.  So, the perturbations sourced by the curvaton will grow as $(\rho_\sigma/\rho_r)^2\propto a^2$ until the curvaton decays.  Thus, PBHs---whose formation depends exponentially on the perturbation amplitude---will preferentially form at the time of the curvaton decay.  Since the scale at which the curvaton decays is unrelated to the scale $k_*$ at which $\varphi$ reaches its minimum, PBHs are expected to form long after the scale at which their primordial overdensities reenter the horizon.\footnote{For example, the authors of Ref.~\cite{Kawasaki:2012wr} considered benchmark points of $M_{\rm min}/M_{\rm BH}=10^{-3}$ and $10^{-8}$.}
But during the intervening time, the (radiation) energy within the comoving volume containing the primordial overdensity will redshift.  The resulting PBH is thus potentially much smaller in mass than if it had formed immediately after horizon reentry.  To produce a black hole of equal mass, $k_*$  must in turn be far smaller than if the PBH collapsed immediately after horizon reentry.  This drives the expected SGW signal to smaller frequencies than can be probed at pulsar timing arrays.  Thus, barring some coincidence between the time it takes $\varphi$ to reach its minimum and the decay time of the curvaton, PTA experiments will not be sensitive to this PBH production model.  Indeed, these time scales are expected to be far apart, with the Hubble parameter at which $\varphi$ oscillations start corresponding to $H \simeq 2 \pi f_\sigma$ and the Hubble parameter when $\varphi$ decays corresponding to $H \simeq \Gamma_\sigma \simeq m_\sigma^3/f_\sigma^2 \simeq \Lambda^6/f_\sigma^5 \ll f_\sigma$, where $\Lambda \ll f_\sigma$ is the explicit $U(1)$ breaking scale where $\sigma$ receives a nonperturbative mass.

{\it Summary:} This model can explain LIGO GWs through merging PBHs, but it has primordial fluctuations sourced after inflation. This generically gives rise to SGWs at frequencies too low for PTAs to discover.

%%%%%%%%%%%%%%%%%%%%%%%%%
\section{Discussion}
\label{sec:discussion}

It is worth noting that we have neglected effects of critical collapse \cite{Gundlach:1999cu,*Gundlach:2002sx,Niemeyer:1997mt,*Niemeyer:1999ak,Kuhnel:2015vtw,Musco:2004ak,*Musco:2008hv,*Musco:2012au}, wherein detailed numerical work has shown that the mass of PBHs formed following horizon reentry may differ from the horizon mass depending on the size of the overdensity in $\delta$.  Most importantly for our discussion, critical collapse effects shift the peak in the mass spectrum to slightly lower masses. This shift in the peak of the mass spectrum can be at most of order a few for the models considered here \cite{Kuhnel:2015vtw}.  To compensate, $\PPhi(k)$ must shift to smaller $k$ (corresponding to larger mass, see Eq.~(\ref{eqn:MPBH})) once critical collapse effects are included in order to keep the PBH abundance spectrum peaked at the same mass.  This change would require a slightly longer collection time, with the collection time depending on the mass as $T \propto \sqrt{M}$.  Of the models we have discussed, the one whose detection prospects are most sensitive to this effect is double inflation with parametric resonance.  In addition, critical collapse may also affect the PBH abundance by an order one factor, which will not significantly impact the SGW abundance.  

We have also neglected effects of nonsphericity \cite{Musco:2012au,Sheth:1999su,Kuhnel:2016exn,Bond:1993we}, which tend to raise the threshold $\delta_c$ on the matter perturbation spectrum for PBHs to form.  For an increase from $\delta_c$ to $\delta_c'$, the amplitude of the scalar spectrum $\PPhi$ must increase by a factor of $\sim (\delta_c'/\delta_c)^2$, and the corresponding GW abundance increases by a factor $\sim (\delta_c'/\delta_c)^4$.  This may increase the SGW abundances in Fig.~\ref{fig:GWspec} by as much as a factor of $\sim 9$.  This is not enough to change the qualitative picture of which models are
probed by PTA experiments, though it may partially compensate for the effects of critical collapse on the needed observation time.

Another effect that can change the necessary threshold on $\delta_c$ is a soft equation of state during the period relevant to PBH formation \cite{Khlopov:1980mg,Jedamzik:1996mr,*Jedamzik:1998hc,Jedamzik:1999am}.  If new physics exists such that the equation of state is more matter-like ($w<1/3$), a smaller $\delta_c$ is necessary to induce collapse.  While for a fixed abundance of PBHs, this can reduce the amplitude of SGWs, it would not change the qualitative picture.

It is worth commenting that the merging of supermassive black holes (SMBH) will also create a stochastic background of gravitational waves.  These obey a power-law spectrum with $\Omega_{\rm GW} \propto f^{2/3}$ \cite{Phinney:2001di,Jaffe:2002rt,Wyithe:2002ep}.  This differs markedly from the spectra considered in this paper, but disentangling the SGW considered here from the GW from the SMBH will present an additional challenge.  This challenge may be acute depending on the amplitude of the SMBH GW.

We now revisit  our assumption that PBHs constitute all of the dark matter.  Recent work \cite{Green:2016xgy} has called into question whether an extended spectrum of PBHs making up all of the DM is allowed by MACHO and faint dwarf cooling constraints.  This, perhaps along with bounds from WMAP \cite{Ricotti:2007au} and Planck \cite{Chen:2016pud}, may indicate that PBHs are allowed to be at most only a fraction of the DM. Nevertheless, even a small abundance of PBHs may still be consistent with LIGO observations \cite{Sasaki:2016jop,Eroshenko:2016hmn}.  But we reiterate that a reduction in the the PBH abundance by even several orders of magnitude will have only a small impact on the SGW spectra considered here because the PBH abundance is exponentially sensitive to the primordial power spectrum amplitude, whereas the SGW abundance is only power-law sensitive to it.  As discussed in the Introduction, the observed LIGO merger rate may be accommodated by PBH DM fractions much smaller than unity.  Since our primary study target is explaining LIGO GWs, not the full abundance of DM, our results hold. 

\section{Conclusion}
\label{sec:conclusion}

If LIGO GWs are to be explained by merging PBHs,
they will need to be produced in the early universe via a peaked primordial curvature spectrum. A byproduct of this will be the production of SGWs.  We have shown that the detectability of SGWs at current and future experiments will depend sensitively on the physics that gives rise to PBHs.  PTAs thus represent a powerful discriminator between PBH production mechanisms with very little dependence on the PBH relic abundance.  Models that give rise to extended initial power spectra are already excluded, while highly peaked models like ones that use parametric resonance give SGW signals detectable by present experiments provided more data collecting time.  
Thus, a future detection of SGWs would give valuable insight into inflation and the formation of PBHs.  Meanwhile, the absence of SGWs at present experiments or even SKA does not rule out every production mechanism for PBHs.  So, if the rate of binary mergers at LIGO continues to agree with the rates calculated in Refs.~\cite{Bird:2016dcv,Sasaki:2016jop,Eroshenko:2016hmn} with increased data collection time, a nondetection of SGWs could indicate a model similar to that of the axion-curvaton as a source of PBHs.  To support the hypothesis that the black holes are primordial and not astrophysical in nature, other probes may be necessary \cite{Clesse:2016vqa,*Clesse:2016ajp,Cholis:2016kqi}.  In any case, inflationary models that explain LIGO GWs through merging PBHs have important implications for SGWs, which will be probed effectively at PTAs.

\acknowledgments{The authors would like to thank Simeon Bird and Julian Mu\~noz for useful discussions.  
This work was supported by the US Department of Energy under grant DE
-SC0007859 and the Humboldt Foundation. 
}

\bibliography{PBH_refs}

\begin{thebibliography}{93}%
\makeatletter
\providecommand \@ifxundefined [1]{%
 \@ifx{#1\undefined}
}%
\providecommand \@ifnum [1]{%
 \ifnum #1\expandafter \@firstoftwo
 \else \expandafter \@secondoftwo
 \fi
}%
\providecommand \@ifx [1]{%
 \ifx #1\expandafter \@firstoftwo
 \else \expandafter \@secondoftwo
 \fi
}%
\providecommand \natexlab [1]{#1}%
\providecommand \enquote  [1]{``#1''}%
\providecommand \bibnamefont  [1]{#1}%
\providecommand \bibfnamefont [1]{#1}%
\providecommand \citenamefont [1]{#1}%
\providecommand \href@noop [0]{\@secondoftwo}%
\providecommand \href [0]{\begingroup \@sanitize@url \@href}%
\providecommand \@href[1]{\@@startlink{#1}\@@href}%
\providecommand \@@href[1]{\endgroup#1\@@endlink}%
\providecommand \@sanitize@url [0]{\catcode `\\12\catcode `\$12\catcode
  `\&12\catcode `\#12\catcode `\^12\catcode `\_12\catcode `\%12\relax}%
\providecommand \@@startlink[1]{}%
\providecommand \@@endlink[0]{}%
\providecommand \url  [0]{\begingroup\@sanitize@url \@url }%
\providecommand \@url [1]{\endgroup\@href {#1}{\urlprefix }}%
\providecommand \urlprefix  [0]{URL }%
\providecommand \Eprint [0]{\href }%
\providecommand \doibase [0]{http://dx.doi.org/}%
\providecommand \selectlanguage [0]{\@gobble}%
\providecommand \bibinfo  [0]{\@secondoftwo}%
\providecommand \bibfield  [0]{\@secondoftwo}%
\providecommand \translation [1]{[#1]}%
\providecommand \BibitemOpen [0]{}%
\providecommand \bibitemStop [0]{}%
\providecommand \bibitemNoStop [0]{.\EOS\space}%
\providecommand \EOS [0]{\spacefactor3000\relax}%
\providecommand \BibitemShut  [1]{\csname bibitem#1\endcsname}%
\let\auto@bib@innerbib\@empty
%</preamble>
\bibitem [{\citenamefont {Abbott}\ \emph
  {et~al.}(2016{\natexlab{a}})\citenamefont {Abbott} \emph
  {et~al.}}]{Abbott:2016blz}%
  \BibitemOpen
  \bibfield  {author} {\bibinfo {author} {\bibfnamefont {B.~P.}\ \bibnamefont
  {Abbott}} \emph {et~al.} (\bibinfo {collaboration} {Virgo, LIGO
  Scientific}),\ }\href {\doibase 10.1103/PhysRevLett.116.061102} {\bibfield
  {journal} {\bibinfo  {journal} {Phys. Rev. Lett.}\ }\textbf {\bibinfo
  {volume} {116}},\ \bibinfo {pages} {061102} (\bibinfo {year}
  {2016}{\natexlab{a}})},\ \Eprint {http://arxiv.org/abs/1602.03837}
  {arXiv:1602.03837 [gr-qc]} \BibitemShut {NoStop}%
%%CITATION = ARXIV:1602.03837;%%
\bibitem [{\citenamefont {Abbott}\ \emph
  {et~al.}(2016{\natexlab{b}})\citenamefont {Abbott} \emph
  {et~al.}}]{Abbott:2016nmj}%
  \BibitemOpen
  \bibfield  {author} {\bibinfo {author} {\bibfnamefont {B.~P.}\ \bibnamefont
  {Abbott}} \emph {et~al.} (\bibinfo {collaboration} {Virgo, LIGO
  Scientific}),\ }\href {\doibase 10.1103/PhysRevLett.116.241103} {\bibfield
  {journal} {\bibinfo  {journal} {Phys. Rev. Lett.}\ }\textbf {\bibinfo
  {volume} {116}},\ \bibinfo {pages} {241103} (\bibinfo {year}
  {2016}{\natexlab{b}})},\ \Eprint {http://arxiv.org/abs/1606.04855}
  {arXiv:1606.04855 [gr-qc]} \BibitemShut {NoStop}%
%%CITATION = ARXIV:1606.04855;%%
\bibitem [{\citenamefont {Carr}\ \emph {et~al.}(2010)\citenamefont {Carr},
  \citenamefont {Kohri}, \citenamefont {Sendouda},\ and\ \citenamefont
  {Yokoyama}}]{Carr:2009jm}%
  \BibitemOpen
  \bibfield  {author} {\bibinfo {author} {\bibfnamefont {B.~J.}\ \bibnamefont
  {Carr}}, \bibinfo {author} {\bibfnamefont {K.}~\bibnamefont {Kohri}},
  \bibinfo {author} {\bibfnamefont {Y.}~\bibnamefont {Sendouda}}, \ and\
  \bibinfo {author} {\bibfnamefont {J.}~\bibnamefont {Yokoyama}},\ }\href
  {\doibase 10.1103/PhysRevD.81.104019} {\bibfield  {journal} {\bibinfo
  {journal} {Phys. Rev.}\ }\textbf {\bibinfo {volume} {D81}},\ \bibinfo {pages}
  {104019} (\bibinfo {year} {2010})},\ \Eprint {http://arxiv.org/abs/0912.5297}
  {arXiv:0912.5297 [astro-ph.CO]} \BibitemShut {NoStop}%
%%CITATION = ARXIV:0912.5297;%%
\bibitem [{\citenamefont {Carr}\ \emph {et~al.}(2016)\citenamefont {Carr},
  \citenamefont {Kuhnel},\ and\ \citenamefont {Sandstad}}]{Carr:2016drx}%
  \BibitemOpen
  \bibfield  {author} {\bibinfo {author} {\bibfnamefont {B.}~\bibnamefont
  {Carr}}, \bibinfo {author} {\bibfnamefont {F.}~\bibnamefont {Kuhnel}}, \ and\
  \bibinfo {author} {\bibfnamefont {M.}~\bibnamefont {Sandstad}},\ }\href
  {\doibase 10.1103/PhysRevD.94.083504} {\bibfield  {journal} {\bibinfo
  {journal} {Phys. Rev.}\ }\textbf {\bibinfo {volume} {D94}},\ \bibinfo {pages}
  {083504} (\bibinfo {year} {2016})},\ \Eprint
  {http://arxiv.org/abs/1607.06077} {arXiv:1607.06077 [astro-ph.CO]}
  \BibitemShut {NoStop}%
%%CITATION = ARXIV:1607.06077;%%
\bibitem [{\citenamefont {Bird}\ \emph {et~al.}(2016)\citenamefont {Bird},
  \citenamefont {Cholis}, \citenamefont {Muñoz}, \citenamefont {Ali-Haïmoud},
  \citenamefont {Kamionkowski}, \citenamefont {Kovetz}, \citenamefont
  {Raccanelli},\ and\ \citenamefont {Riess}}]{Bird:2016dcv}%
  \BibitemOpen
  \bibfield  {author} {\bibinfo {author} {\bibfnamefont {S.}~\bibnamefont
  {Bird}}, \bibinfo {author} {\bibfnamefont {I.}~\bibnamefont {Cholis}},
  \bibinfo {author} {\bibfnamefont {J.~B.}\ \bibnamefont {Muñoz}}, \bibinfo
  {author} {\bibfnamefont {Y.}~\bibnamefont {Ali-Haïmoud}}, \bibinfo {author}
  {\bibfnamefont {M.}~\bibnamefont {Kamionkowski}}, \bibinfo {author}
  {\bibfnamefont {E.~D.}\ \bibnamefont {Kovetz}}, \bibinfo {author}
  {\bibfnamefont {A.}~\bibnamefont {Raccanelli}}, \ and\ \bibinfo {author}
  {\bibfnamefont {A.~G.}\ \bibnamefont {Riess}},\ }\href {\doibase
  10.1103/PhysRevLett.116.201301} {\bibfield  {journal} {\bibinfo  {journal}
  {Phys. Rev. Lett.}\ }\textbf {\bibinfo {volume} {116}},\ \bibinfo {pages}
  {201301} (\bibinfo {year} {2016})},\ \Eprint
  {http://arxiv.org/abs/1603.00464} {arXiv:1603.00464 [astro-ph.CO]}
  \BibitemShut {NoStop}%
%%CITATION = ARXIV:1603.00464;%%
\bibitem [{\citenamefont {Sasaki}\ \emph {et~al.}(2016)\citenamefont {Sasaki},
  \citenamefont {Suyama}, \citenamefont {Tanaka},\ and\ \citenamefont
  {Yokoyama}}]{Sasaki:2016jop}%
  \BibitemOpen
  \bibfield  {author} {\bibinfo {author} {\bibfnamefont {M.}~\bibnamefont
  {Sasaki}}, \bibinfo {author} {\bibfnamefont {T.}~\bibnamefont {Suyama}},
  \bibinfo {author} {\bibfnamefont {T.}~\bibnamefont {Tanaka}}, \ and\ \bibinfo
  {author} {\bibfnamefont {S.}~\bibnamefont {Yokoyama}},\ }\href {\doibase
  10.1103/PhysRevLett.117.061101} {\bibfield  {journal} {\bibinfo  {journal}
  {Phys. Rev. Lett.}\ }\textbf {\bibinfo {volume} {117}},\ \bibinfo {pages}
  {061101} (\bibinfo {year} {2016})},\ \Eprint
  {http://arxiv.org/abs/1603.08338} {arXiv:1603.08338 [astro-ph.CO]}
  \BibitemShut {NoStop}%
%%CITATION = ARXIV:1603.08338;%%
\bibitem [{\citenamefont {Eroshenko}(2016)}]{Eroshenko:2016hmn}%
  \BibitemOpen
  \bibfield  {author} {\bibinfo {author} {\bibfnamefont {{\relax Yu}.~N.}\
  \bibnamefont {Eroshenko}},\ }\href@noop {} {\  (\bibinfo {year} {2016})},\
  \Eprint {http://arxiv.org/abs/1604.04932} {arXiv:1604.04932 [astro-ph.CO]}
  \BibitemShut {NoStop}%
%%CITATION = ARXIV:1604.04932;%%
\bibitem [{\citenamefont {Tisserand}\ \emph {et~al.}(2007)\citenamefont
  {Tisserand} \emph {et~al.}}]{Tisserand:2006zx}%
  \BibitemOpen
  \bibfield  {author} {\bibinfo {author} {\bibfnamefont {P.}~\bibnamefont
  {Tisserand}} \emph {et~al.} (\bibinfo {collaboration} {EROS-2}),\ }\href
  {\doibase 10.1051/0004-6361:20066017} {\bibfield  {journal} {\bibinfo
  {journal} {Astron. Astrophys.}\ }\textbf {\bibinfo {volume} {469}},\ \bibinfo
  {pages} {387} (\bibinfo {year} {2007})},\ \Eprint
  {http://arxiv.org/abs/astro-ph/0607207} {arXiv:astro-ph/0607207 [astro-ph]}
  \BibitemShut {NoStop}%
%%CITATION = ASTRO-PH/0607207;%%
\bibitem [{\citenamefont {Wyrzykowski}\ \emph {et~al.}(2011)\citenamefont
  {Wyrzykowski} \emph {et~al.}}]{Wyrzykowski:2011tr}%
  \BibitemOpen
  \bibfield  {author} {\bibinfo {author} {\bibfnamefont {L.}~\bibnamefont
  {Wyrzykowski}} \emph {et~al.},\ }\href {\doibase
  10.1111/j.1365-2966.2011.19243.x} {\bibfield  {journal} {\bibinfo  {journal}
  {Mon. Not. Roy. Astron. Soc.}\ }\textbf {\bibinfo {volume} {416}},\ \bibinfo
  {pages} {2949} (\bibinfo {year} {2011})},\ \Eprint
  {http://arxiv.org/abs/1106.2925} {arXiv:1106.2925 [astro-ph.GA]} \BibitemShut
  {NoStop}%
%%CITATION = ARXIV:1106.2925;%%
\bibitem [{\citenamefont {Allsman}\ \emph {et~al.}(2001)\citenamefont {Allsman}
  \emph {et~al.}}]{Allsman:2000kg}%
  \BibitemOpen
  \bibfield  {author} {\bibinfo {author} {\bibfnamefont {R.~A.}\ \bibnamefont
  {Allsman}} \emph {et~al.} (\bibinfo {collaboration} {Macho}),\ }\href
  {\doibase 10.1086/319636} {\bibfield  {journal} {\bibinfo  {journal}
  {Astrophys. J.}\ }\textbf {\bibinfo {volume} {550}},\ \bibinfo {pages} {L169}
  (\bibinfo {year} {2001})},\ \Eprint {http://arxiv.org/abs/astro-ph/0011506}
  {arXiv:astro-ph/0011506 [astro-ph]} \BibitemShut {NoStop}%
%%CITATION = ASTRO-PH/0011506;%%
\bibitem [{\citenamefont {Brandt}(2016)}]{Brandt:2016aco}%
  \BibitemOpen
  \bibfield  {author} {\bibinfo {author} {\bibfnamefont {T.~D.}\ \bibnamefont
  {Brandt}},\ }\href {\doibase 10.3847/2041-8205/824/2/L31} {\bibfield
  {journal} {\bibinfo  {journal} {Astrophys. J.}\ }\textbf {\bibinfo {volume}
  {824}},\ \bibinfo {pages} {L31} (\bibinfo {year} {2016})},\ \Eprint
  {http://arxiv.org/abs/1605.03665} {arXiv:1605.03665 [astro-ph.GA]}
  \BibitemShut {NoStop}%
%%CITATION = ARXIV:1605.03665;%%
\bibitem [{\citenamefont {Green}(2016)}]{Green:2016xgy}%
  \BibitemOpen
  \bibfield  {author} {\bibinfo {author} {\bibfnamefont {A.~M.}\ \bibnamefont
  {Green}},\ }\href {\doibase 10.1103/PhysRevD.94.063530} {\bibfield  {journal}
  {\bibinfo  {journal} {Phys. Rev.}\ }\textbf {\bibinfo {volume} {D94}},\
  \bibinfo {pages} {063530} (\bibinfo {year} {2016})},\ \Eprint
  {http://arxiv.org/abs/1609.01143} {arXiv:1609.01143 [astro-ph.CO]}
  \BibitemShut {NoStop}%
%%CITATION = ARXIV:1609.01143;%%
\bibitem [{\citenamefont {Schutz}\ and\ \citenamefont
  {Liu}(2017)}]{Schutz:2016khr}%
  \BibitemOpen
  \bibfield  {author} {\bibinfo {author} {\bibfnamefont {K.}~\bibnamefont
  {Schutz}}\ and\ \bibinfo {author} {\bibfnamefont {A.}~\bibnamefont {Liu}},\
  }\href {\doibase 10.1103/PhysRevD.95.023002} {\bibfield  {journal} {\bibinfo
  {journal} {Phys. Rev.}\ }\textbf {\bibinfo {volume} {D95}},\ \bibinfo {pages}
  {023002} (\bibinfo {year} {2017})},\ \Eprint
  {http://arxiv.org/abs/1610.04234} {arXiv:1610.04234 [astro-ph.CO]}
  \BibitemShut {NoStop}%
%%CITATION = ARXIV:1610.04234;%%
\bibitem [{\citenamefont {Ricotti}\ \emph {et~al.}(2008)\citenamefont
  {Ricotti}, \citenamefont {Ostriker},\ and\ \citenamefont
  {Mack}}]{Ricotti:2007au}%
  \BibitemOpen
  \bibfield  {author} {\bibinfo {author} {\bibfnamefont {M.}~\bibnamefont
  {Ricotti}}, \bibinfo {author} {\bibfnamefont {J.~P.}\ \bibnamefont
  {Ostriker}}, \ and\ \bibinfo {author} {\bibfnamefont {K.~J.}\ \bibnamefont
  {Mack}},\ }\href {\doibase 10.1086/587831} {\bibfield  {journal} {\bibinfo
  {journal} {Astrophys. J.}\ }\textbf {\bibinfo {volume} {680}},\ \bibinfo
  {pages} {829} (\bibinfo {year} {2008})},\ \Eprint
  {http://arxiv.org/abs/0709.0524} {arXiv:0709.0524 [astro-ph]} \BibitemShut
  {NoStop}%
%%CITATION = ARXIV:0709.0524;%%
\bibitem [{\citenamefont {Chen}\ \emph {et~al.}(2016)\citenamefont {Chen},
  \citenamefont {Huang},\ and\ \citenamefont {Wang}}]{Chen:2016pud}%
  \BibitemOpen
  \bibfield  {author} {\bibinfo {author} {\bibfnamefont {L.}~\bibnamefont
  {Chen}}, \bibinfo {author} {\bibfnamefont {Q.-G.}\ \bibnamefont {Huang}}, \
  and\ \bibinfo {author} {\bibfnamefont {K.}~\bibnamefont {Wang}},\ }\href
  {\doibase 10.1088/1475-7516/2016/12/044} {\bibfield  {journal} {\bibinfo
  {journal} {JCAP}\ }\textbf {\bibinfo {volume} {1612}},\ \bibinfo {pages}
  {044} (\bibinfo {year} {2016})},\ \Eprint {http://arxiv.org/abs/1608.02174}
  {arXiv:1608.02174 [astro-ph.CO]} \BibitemShut {NoStop}%
%%CITATION = ARXIV:1608.02174;%%
\bibitem [{\citenamefont {Nakamura}\ \emph {et~al.}(1997)\citenamefont
  {Nakamura}, \citenamefont {Sasaki}, \citenamefont {Tanaka},\ and\
  \citenamefont {Thorne}}]{Nakamura:1997sm}%
  \BibitemOpen
  \bibfield  {author} {\bibinfo {author} {\bibfnamefont {T.}~\bibnamefont
  {Nakamura}}, \bibinfo {author} {\bibfnamefont {M.}~\bibnamefont {Sasaki}},
  \bibinfo {author} {\bibfnamefont {T.}~\bibnamefont {Tanaka}}, \ and\ \bibinfo
  {author} {\bibfnamefont {K.~S.}\ \bibnamefont {Thorne}},\ }\href {\doibase
  10.1086/310886} {\bibfield  {journal} {\bibinfo  {journal} {Astrophys. J.}\
  }\textbf {\bibinfo {volume} {487}},\ \bibinfo {pages} {L139} (\bibinfo {year}
  {1997})},\ \Eprint {http://arxiv.org/abs/astro-ph/9708060}
  {arXiv:astro-ph/9708060 [astro-ph]} \BibitemShut {NoStop}%
%%CITATION = ASTRO-PH/9708060;%%
\bibitem [{\citenamefont {Ioka}\ \emph {et~al.}(1998)\citenamefont {Ioka},
  \citenamefont {Chiba}, \citenamefont {Tanaka},\ and\ \citenamefont
  {Nakamura}}]{Ioka:1998nz}%
  \BibitemOpen
  \bibfield  {author} {\bibinfo {author} {\bibfnamefont {K.}~\bibnamefont
  {Ioka}}, \bibinfo {author} {\bibfnamefont {T.}~\bibnamefont {Chiba}},
  \bibinfo {author} {\bibfnamefont {T.}~\bibnamefont {Tanaka}}, \ and\ \bibinfo
  {author} {\bibfnamefont {T.}~\bibnamefont {Nakamura}},\ }\href {\doibase
  10.1103/PhysRevD.58.063003} {\bibfield  {journal} {\bibinfo  {journal} {Phys.
  Rev.}\ }\textbf {\bibinfo {volume} {D58}},\ \bibinfo {pages} {063003}
  (\bibinfo {year} {1998})},\ \Eprint {http://arxiv.org/abs/astro-ph/9807018}
  {arXiv:astro-ph/9807018 [astro-ph]} \BibitemShut {NoStop}%
%%CITATION = ASTRO-PH/9807018;%%
\bibitem [{\citenamefont {Tomita}(1967)}]{Tomita:1967}%
  \BibitemOpen
  \bibfield  {author} {\bibinfo {author} {\bibfnamefont {K.}~\bibnamefont
  {Tomita}},\ }\href {\doibase 10.1143/PTP.37.831} {\bibfield  {journal}
  {\bibinfo  {journal} {Prog. Theor. Phys.}\ }\textbf {\bibinfo {volume}
  {37}},\ \bibinfo {pages} {831} (\bibinfo {year} {1967})}\BibitemShut
  {NoStop}%
\bibitem [{\citenamefont {Matarrese}\ \emph {et~al.}(1993)\citenamefont
  {Matarrese}, \citenamefont {Pantano},\ and\ \citenamefont
  {Saez}}]{Matarrese:1992rp}%
  \BibitemOpen
  \bibfield  {author} {\bibinfo {author} {\bibfnamefont {S.}~\bibnamefont
  {Matarrese}}, \bibinfo {author} {\bibfnamefont {O.}~\bibnamefont {Pantano}},
  \ and\ \bibinfo {author} {\bibfnamefont {D.}~\bibnamefont {Saez}},\ }\href
  {\doibase 10.1103/PhysRevD.47.1311} {\bibfield  {journal} {\bibinfo
  {journal} {Phys. Rev.}\ }\textbf {\bibinfo {volume} {D47}},\ \bibinfo {pages}
  {1311} (\bibinfo {year} {1993})}\BibitemShut {NoStop}%
%%CITATION = PHRVA,D47,1311;%%
\bibitem [{\citenamefont {Matarrese}\ \emph {et~al.}(1994)\citenamefont
  {Matarrese}, \citenamefont {Pantano},\ and\ \citenamefont
  {Saez}}]{Matarrese:1993zf}%
  \BibitemOpen
  \bibfield  {author} {\bibinfo {author} {\bibfnamefont {S.}~\bibnamefont
  {Matarrese}}, \bibinfo {author} {\bibfnamefont {O.}~\bibnamefont {Pantano}},
  \ and\ \bibinfo {author} {\bibfnamefont {D.}~\bibnamefont {Saez}},\ }\href
  {\doibase 10.1103/PhysRevLett.72.320} {\bibfield  {journal} {\bibinfo
  {journal} {Phys. Rev. Lett.}\ }\textbf {\bibinfo {volume} {72}},\ \bibinfo
  {pages} {320} (\bibinfo {year} {1994})},\ \Eprint
  {http://arxiv.org/abs/astro-ph/9310036} {arXiv:astro-ph/9310036 [astro-ph]}
  \BibitemShut {NoStop}%
%%CITATION = ASTRO-PH/9310036;%%
\bibitem [{\citenamefont {Matarrese}\ \emph {et~al.}(1998)\citenamefont
  {Matarrese}, \citenamefont {Mollerach},\ and\ \citenamefont
  {Bruni}}]{Matarrese:1997ay}%
  \BibitemOpen
  \bibfield  {author} {\bibinfo {author} {\bibfnamefont {S.}~\bibnamefont
  {Matarrese}}, \bibinfo {author} {\bibfnamefont {S.}~\bibnamefont
  {Mollerach}}, \ and\ \bibinfo {author} {\bibfnamefont {M.}~\bibnamefont
  {Bruni}},\ }\href {\doibase 10.1103/PhysRevD.58.043504} {\bibfield  {journal}
  {\bibinfo  {journal} {Phys. Rev.}\ }\textbf {\bibinfo {volume} {D58}},\
  \bibinfo {pages} {043504} (\bibinfo {year} {1998})},\ \Eprint
  {http://arxiv.org/abs/astro-ph/9707278} {arXiv:astro-ph/9707278 [astro-ph]}
  \BibitemShut {NoStop}%
%%CITATION = ASTRO-PH/9707278;%%
\bibitem [{\citenamefont {Noh}\ and\ \citenamefont {Hwang}(2004)}]{Noh:2004bc}%
  \BibitemOpen
  \bibfield  {author} {\bibinfo {author} {\bibfnamefont {H.}~\bibnamefont
  {Noh}}\ and\ \bibinfo {author} {\bibfnamefont {J.-c.}\ \bibnamefont
  {Hwang}},\ }\href {\doibase 10.1103/PhysRevD.69.104011} {\bibfield  {journal}
  {\bibinfo  {journal} {Phys. Rev.}\ }\textbf {\bibinfo {volume} {D69}},\
  \bibinfo {pages} {104011} (\bibinfo {year} {2004})}\BibitemShut {NoStop}%
%%CITATION = PHRVA,D69,104011;%%
\bibitem [{\citenamefont {Carbone}\ and\ \citenamefont
  {Matarrese}(2005)}]{Carbone:2004iv}%
  \BibitemOpen
  \bibfield  {author} {\bibinfo {author} {\bibfnamefont {C.}~\bibnamefont
  {Carbone}}\ and\ \bibinfo {author} {\bibfnamefont {S.}~\bibnamefont
  {Matarrese}},\ }\href {\doibase 10.1103/PhysRevD.71.043508} {\bibfield
  {journal} {\bibinfo  {journal} {Phys. Rev.}\ }\textbf {\bibinfo {volume}
  {D71}},\ \bibinfo {pages} {043508} (\bibinfo {year} {2005})},\ \Eprint
  {http://arxiv.org/abs/astro-ph/0407611} {arXiv:astro-ph/0407611 [astro-ph]}
  \BibitemShut {NoStop}%
%%CITATION = ASTRO-PH/0407611;%%
\bibitem [{\citenamefont {Detweiler}(1979)}]{Detweiler:1979wn}%
  \BibitemOpen
  \bibfield  {author} {\bibinfo {author} {\bibfnamefont {S.~L.}\ \bibnamefont
  {Detweiler}},\ }\href {\doibase 10.1086/157593} {\bibfield  {journal}
  {\bibinfo  {journal} {Astrophys. J.}\ }\textbf {\bibinfo {volume} {234}},\
  \bibinfo {pages} {1100} (\bibinfo {year} {1979})}\BibitemShut {NoStop}%
%%CITATION = ASJOA,234,1100;%%
\bibitem [{\citenamefont {Saito}\ and\ \citenamefont
  {Yokoyama}(2009)}]{Saito:2008jc}%
  \BibitemOpen
  \bibfield  {author} {\bibinfo {author} {\bibfnamefont {R.}~\bibnamefont
  {Saito}}\ and\ \bibinfo {author} {\bibfnamefont {J.}~\bibnamefont
  {Yokoyama}},\ }\href {\doibase 10.1103/PhysRevLett.102.161101,
  10.1103/PhysRevLett.107.069901} {\bibfield  {journal} {\bibinfo  {journal}
  {Phys. Rev. Lett.}\ }\textbf {\bibinfo {volume} {102}},\ \bibinfo {pages}
  {161101} (\bibinfo {year} {2009})},\ \bibinfo {note} {[Erratum: Phys. Rev.
  Lett.107,069901(2011)]},\ \Eprint {http://arxiv.org/abs/0812.4339}
  {arXiv:0812.4339 [astro-ph]} \BibitemShut {NoStop}%
%%CITATION = ARXIV:0812.4339;%%
\bibitem [{\citenamefont {Saito}\ and\ \citenamefont
  {Yokoyama}(2010)}]{Saito:2009jt}%
  \BibitemOpen
  \bibfield  {author} {\bibinfo {author} {\bibfnamefont {R.}~\bibnamefont
  {Saito}}\ and\ \bibinfo {author} {\bibfnamefont {J.}~\bibnamefont
  {Yokoyama}},\ }\href {\doibase 10.1143/PTP.126.351, 10.1143/PTP.123.867}
  {\bibfield  {journal} {\bibinfo  {journal} {. Theor. Phys.}\ }\textbf
  {\bibinfo {volume} {123}},\ \bibinfo {pages} {867} (\bibinfo {year}
  {2010})},\ \bibinfo {note} {[Erratum: Prog. Theor. Phys.126,351(2011)]},\
  \Eprint {http://arxiv.org/abs/0912.5317} {arXiv:0912.5317 [astro-ph.CO]}
  \BibitemShut {NoStop}%
%%CITATION = ARXIV:0912.5317;%%
\bibitem [{\citenamefont {Bugaev}\ and\ \citenamefont
  {Klimai}(2011)}]{Bugaev:2010bb}%
  \BibitemOpen
  \bibfield  {author} {\bibinfo {author} {\bibfnamefont {E.}~\bibnamefont
  {Bugaev}}\ and\ \bibinfo {author} {\bibfnamefont {P.}~\bibnamefont
  {Klimai}},\ }\href {\doibase 10.1103/PhysRevD.83.083521} {\bibfield
  {journal} {\bibinfo  {journal} {Phys. Rev.}\ }\textbf {\bibinfo {volume}
  {D83}},\ \bibinfo {pages} {083521} (\bibinfo {year} {2011})},\ \Eprint
  {http://arxiv.org/abs/1012.4697} {arXiv:1012.4697 [astro-ph.CO]} \BibitemShut
  {NoStop}%
%%CITATION = ARXIV:1012.4697;%%
\bibitem [{\citenamefont {Garcia-Bellido}\ \emph {et~al.}(2016)\citenamefont
  {Garcia-Bellido}, \citenamefont {Peloso},\ and\ \citenamefont
  {Unal}}]{Garcia-Bellido:2016dkw}%
  \BibitemOpen
  \bibfield  {author} {\bibinfo {author} {\bibfnamefont {J.}~\bibnamefont
  {Garcia-Bellido}}, \bibinfo {author} {\bibfnamefont {M.}~\bibnamefont
  {Peloso}}, \ and\ \bibinfo {author} {\bibfnamefont {C.}~\bibnamefont
  {Unal}},\ }\href {\doibase 10.1088/1475-7516/2016/12/031} {\bibfield
  {journal} {\bibinfo  {journal} {JCAP}\ }\textbf {\bibinfo {volume} {1612}},\
  \bibinfo {pages} {031} (\bibinfo {year} {2016})},\ \Eprint
  {http://arxiv.org/abs/1610.03763} {arXiv:1610.03763 [astro-ph.CO]}
  \BibitemShut {NoStop}%
%%CITATION = ARXIV:1610.03763;%%
\bibitem [{\citenamefont {Inomata}\ \emph {et~al.}(2016)\citenamefont
  {Inomata}, \citenamefont {Kawasaki}, \citenamefont {Mukaida}, \citenamefont
  {Tada},\ and\ \citenamefont {Yanagida}}]{Inomata:2016rbd}%
  \BibitemOpen
  \bibfield  {author} {\bibinfo {author} {\bibfnamefont {K.}~\bibnamefont
  {Inomata}}, \bibinfo {author} {\bibfnamefont {M.}~\bibnamefont {Kawasaki}},
  \bibinfo {author} {\bibfnamefont {K.}~\bibnamefont {Mukaida}}, \bibinfo
  {author} {\bibfnamefont {Y.}~\bibnamefont {Tada}}, \ and\ \bibinfo {author}
  {\bibfnamefont {T.~T.}\ \bibnamefont {Yanagida}},\ }\href@noop {} {\
  (\bibinfo {year} {2016})},\ \Eprint {http://arxiv.org/abs/1611.06130}
  {arXiv:1611.06130 [astro-ph.CO]} \BibitemShut {NoStop}%
%%CITATION = ARXIV:1611.06130;%%
\bibitem [{\citenamefont {Josan}\ \emph {et~al.}(2009)\citenamefont {Josan},
  \citenamefont {Green},\ and\ \citenamefont {Malik}}]{Josan:2009qn}%
  \BibitemOpen
  \bibfield  {author} {\bibinfo {author} {\bibfnamefont {A.~S.}\ \bibnamefont
  {Josan}}, \bibinfo {author} {\bibfnamefont {A.~M.}\ \bibnamefont {Green}}, \
  and\ \bibinfo {author} {\bibfnamefont {K.~A.}\ \bibnamefont {Malik}},\ }\href
  {\doibase 10.1103/PhysRevD.79.103520} {\bibfield  {journal} {\bibinfo
  {journal} {Phys. Rev.}\ }\textbf {\bibinfo {volume} {D79}},\ \bibinfo {pages}
  {103520} (\bibinfo {year} {2009})},\ \Eprint {http://arxiv.org/abs/0903.3184}
  {arXiv:0903.3184 [astro-ph.CO]} \BibitemShut {NoStop}%
%%CITATION = ARXIV:0903.3184;%%
\bibitem [{\citenamefont {Press}\ and\ \citenamefont
  {Schechter}(1974)}]{Press:1973iz}%
  \BibitemOpen
  \bibfield  {author} {\bibinfo {author} {\bibfnamefont {W.~H.}\ \bibnamefont
  {Press}}\ and\ \bibinfo {author} {\bibfnamefont {P.}~\bibnamefont
  {Schechter}},\ }\href {\doibase 10.1086/152650} {\bibfield  {journal}
  {\bibinfo  {journal} {Astrophys. J.}\ }\textbf {\bibinfo {volume} {187}},\
  \bibinfo {pages} {425} (\bibinfo {year} {1974})}\BibitemShut {NoStop}%
%%CITATION = ASJOA,187,425;%%
\bibitem [{\citenamefont {Kawasaki}\ \emph {et~al.}(2016)\citenamefont
  {Kawasaki}, \citenamefont {Kusenko}, \citenamefont {Tada},\ and\
  \citenamefont {Yanagida}}]{Kawasaki:2016pql}%
  \BibitemOpen
  \bibfield  {author} {\bibinfo {author} {\bibfnamefont {M.}~\bibnamefont
  {Kawasaki}}, \bibinfo {author} {\bibfnamefont {A.}~\bibnamefont {Kusenko}},
  \bibinfo {author} {\bibfnamefont {Y.}~\bibnamefont {Tada}}, \ and\ \bibinfo
  {author} {\bibfnamefont {T.~T.}\ \bibnamefont {Yanagida}},\ }\href {\doibase
  10.1103/PhysRevD.94.083523} {\bibfield  {journal} {\bibinfo  {journal} {Phys.
  Rev.}\ }\textbf {\bibinfo {volume} {D94}},\ \bibinfo {pages} {083523}
  (\bibinfo {year} {2016})},\ \Eprint {http://arxiv.org/abs/1606.07631}
  {arXiv:1606.07631 [astro-ph.CO]} \BibitemShut {NoStop}%
%%CITATION = ARXIV:1606.07631;%%
\bibitem [{\citenamefont {Young}\ \emph {et~al.}(2014)\citenamefont {Young},
  \citenamefont {Byrnes},\ and\ \citenamefont {Sasaki}}]{Young:2014ana}%
  \BibitemOpen
  \bibfield  {author} {\bibinfo {author} {\bibfnamefont {S.}~\bibnamefont
  {Young}}, \bibinfo {author} {\bibfnamefont {C.~T.}\ \bibnamefont {Byrnes}}, \
  and\ \bibinfo {author} {\bibfnamefont {M.}~\bibnamefont {Sasaki}},\ }\href
  {\doibase 10.1088/1475-7516/2014/07/045} {\bibfield  {journal} {\bibinfo
  {journal} {JCAP}\ }\textbf {\bibinfo {volume} {1407}},\ \bibinfo {pages}
  {045} (\bibinfo {year} {2014})},\ \Eprint {http://arxiv.org/abs/1405.7023}
  {arXiv:1405.7023 [gr-qc]} \BibitemShut {NoStop}%
%%CITATION = ARXIV:1405.7023;%%
\bibitem [{\citenamefont {Carr}(1975)}]{Carr:1975qj}%
  \BibitemOpen
  \bibfield  {author} {\bibinfo {author} {\bibfnamefont {B.~J.}\ \bibnamefont
  {Carr}},\ }\href {\doibase 10.1086/153853} {\bibfield  {journal} {\bibinfo
  {journal} {Astrophys. J.}\ }\textbf {\bibinfo {volume} {201}},\ \bibinfo
  {pages} {1} (\bibinfo {year} {1975})}\BibitemShut {NoStop}%
%%CITATION = ASJOA,201,1;%%
\bibitem [{\citenamefont {Musco}\ \emph {et~al.}(2005)\citenamefont {Musco},
  \citenamefont {Miller},\ and\ \citenamefont {Rezzolla}}]{Musco:2004ak}%
  \BibitemOpen
  \bibfield  {author} {\bibinfo {author} {\bibfnamefont {I.}~\bibnamefont
  {Musco}}, \bibinfo {author} {\bibfnamefont {J.~C.}\ \bibnamefont {Miller}}, \
  and\ \bibinfo {author} {\bibfnamefont {L.}~\bibnamefont {Rezzolla}},\ }\href
  {\doibase 10.1088/0264-9381/22/7/013} {\bibfield  {journal} {\bibinfo
  {journal} {Class. Quant. Grav.}\ }\textbf {\bibinfo {volume} {22}},\ \bibinfo
  {pages} {1405} (\bibinfo {year} {2005})},\ \Eprint
  {http://arxiv.org/abs/gr-qc/0412063} {arXiv:gr-qc/0412063 [gr-qc]}
  \BibitemShut {NoStop}%
%%CITATION = GR-QC/0412063;%%
\bibitem [{\citenamefont {Musco}\ \emph {et~al.}(2009)\citenamefont {Musco},
  \citenamefont {Miller},\ and\ \citenamefont {Polnarev}}]{Musco:2008hv}%
  \BibitemOpen
  \bibfield  {author} {\bibinfo {author} {\bibfnamefont {I.}~\bibnamefont
  {Musco}}, \bibinfo {author} {\bibfnamefont {J.~C.}\ \bibnamefont {Miller}}, \
  and\ \bibinfo {author} {\bibfnamefont {A.~G.}\ \bibnamefont {Polnarev}},\
  }\href {\doibase 10.1088/0264-9381/26/23/235001} {\bibfield  {journal}
  {\bibinfo  {journal} {Class. Quant. Grav.}\ }\textbf {\bibinfo {volume}
  {26}},\ \bibinfo {pages} {235001} (\bibinfo {year} {2009})},\ \Eprint
  {http://arxiv.org/abs/0811.1452} {arXiv:0811.1452 [gr-qc]} \BibitemShut
  {NoStop}%
%%CITATION = ARXIV:0811.1452;%%
\bibitem [{\citenamefont {Musco}\ and\ \citenamefont
  {Miller}(2013)}]{Musco:2012au}%
  \BibitemOpen
  \bibfield  {author} {\bibinfo {author} {\bibfnamefont {I.}~\bibnamefont
  {Musco}}\ and\ \bibinfo {author} {\bibfnamefont {J.~C.}\ \bibnamefont
  {Miller}},\ }\href {\doibase 10.1088/0264-9381/30/14/145009} {\bibfield
  {journal} {\bibinfo  {journal} {Class. Quant. Grav.}\ }\textbf {\bibinfo
  {volume} {30}},\ \bibinfo {pages} {145009} (\bibinfo {year} {2013})},\
  \Eprint {http://arxiv.org/abs/1201.2379} {arXiv:1201.2379 [gr-qc]}
  \BibitemShut {NoStop}%
%%CITATION = ARXIV:1201.2379;%%
\bibitem [{\citenamefont {Sheth}\ \emph {et~al.}(2001)\citenamefont {Sheth},
  \citenamefont {Mo},\ and\ \citenamefont {Tormen}}]{Sheth:1999su}%
  \BibitemOpen
  \bibfield  {author} {\bibinfo {author} {\bibfnamefont {R.~K.}\ \bibnamefont
  {Sheth}}, \bibinfo {author} {\bibfnamefont {H.~J.}\ \bibnamefont {Mo}}, \
  and\ \bibinfo {author} {\bibfnamefont {G.}~\bibnamefont {Tormen}},\ }\href
  {\doibase 10.1046/j.1365-8711.2001.04006.x} {\bibfield  {journal} {\bibinfo
  {journal} {Mon. Not. Roy. Astron. Soc.}\ }\textbf {\bibinfo {volume} {323}},\
  \bibinfo {pages} {1} (\bibinfo {year} {2001})},\ \Eprint
  {http://arxiv.org/abs/astro-ph/9907024} {arXiv:astro-ph/9907024 [astro-ph]}
  \BibitemShut {NoStop}%
%%CITATION = ASTRO-PH/9907024;%%
\bibitem [{\citenamefont {Kuhnel}\ and\ \citenamefont
  {Sandstad}(2016)}]{Kuhnel:2016exn}%
  \BibitemOpen
  \bibfield  {author} {\bibinfo {author} {\bibfnamefont {F.}~\bibnamefont
  {Kuhnel}}\ and\ \bibinfo {author} {\bibfnamefont {M.}~\bibnamefont
  {Sandstad}},\ }\href {\doibase 10.1103/PhysRevD.94.063514} {\bibfield
  {journal} {\bibinfo  {journal} {Phys. Rev.}\ }\textbf {\bibinfo {volume}
  {D94}},\ \bibinfo {pages} {063514} (\bibinfo {year} {2016})},\ \Eprint
  {http://arxiv.org/abs/1602.04815} {arXiv:1602.04815 [astro-ph.CO]}
  \BibitemShut {NoStop}%
%%CITATION = ARXIV:1602.04815;%%
\bibitem [{\citenamefont {Bond}\ and\ \citenamefont
  {Myers}(1996)}]{Bond:1993we}%
  \BibitemOpen
  \bibfield  {author} {\bibinfo {author} {\bibfnamefont {J.~R.}\ \bibnamefont
  {Bond}}\ and\ \bibinfo {author} {\bibfnamefont {S.~T.}\ \bibnamefont
  {Myers}},\ }\href {\doibase 10.1086/192267} {\bibfield  {journal} {\bibinfo
  {journal} {Astrophys. J. Suppl.}\ }\textbf {\bibinfo {volume} {103}},\
  \bibinfo {pages} {1} (\bibinfo {year} {1996})}\BibitemShut {NoStop}%
%%CITATION = APJSA,103,1;%%
\bibitem [{\citenamefont {Ananda}\ \emph {et~al.}(2007)\citenamefont {Ananda},
  \citenamefont {Clarkson},\ and\ \citenamefont {Wands}}]{Ananda:2006af}%
  \BibitemOpen
  \bibfield  {author} {\bibinfo {author} {\bibfnamefont {K.~N.}\ \bibnamefont
  {Ananda}}, \bibinfo {author} {\bibfnamefont {C.}~\bibnamefont {Clarkson}}, \
  and\ \bibinfo {author} {\bibfnamefont {D.}~\bibnamefont {Wands}},\ }\href
  {\doibase 10.1103/PhysRevD.75.123518} {\bibfield  {journal} {\bibinfo
  {journal} {Phys. Rev.}\ }\textbf {\bibinfo {volume} {D75}},\ \bibinfo {pages}
  {123518} (\bibinfo {year} {2007})},\ \Eprint
  {http://arxiv.org/abs/gr-qc/0612013} {arXiv:gr-qc/0612013 [gr-qc]}
  \BibitemShut {NoStop}%
%%CITATION = GR-QC/0612013;%%
\bibitem [{\citenamefont {Baumann}\ \emph {et~al.}(2007)\citenamefont
  {Baumann}, \citenamefont {Steinhardt}, \citenamefont {Takahashi},\ and\
  \citenamefont {Ichiki}}]{Baumann:2007zm}%
  \BibitemOpen
  \bibfield  {author} {\bibinfo {author} {\bibfnamefont {D.}~\bibnamefont
  {Baumann}}, \bibinfo {author} {\bibfnamefont {P.~J.}\ \bibnamefont
  {Steinhardt}}, \bibinfo {author} {\bibfnamefont {K.}~\bibnamefont
  {Takahashi}}, \ and\ \bibinfo {author} {\bibfnamefont {K.}~\bibnamefont
  {Ichiki}},\ }\href {\doibase 10.1103/PhysRevD.76.084019} {\bibfield
  {journal} {\bibinfo  {journal} {Phys. Rev.}\ }\textbf {\bibinfo {volume}
  {D76}},\ \bibinfo {pages} {084019} (\bibinfo {year} {2007})},\ \Eprint
  {http://arxiv.org/abs/hep-th/0703290} {arXiv:hep-th/0703290 [hep-th]}
  \BibitemShut {NoStop}%
%%CITATION = HEP-TH/0703290;%%
\bibitem [{\citenamefont {Bugaev}\ and\ \citenamefont
  {Klimai}(2010)}]{Bugaev:2009zh}%
  \BibitemOpen
  \bibfield  {author} {\bibinfo {author} {\bibfnamefont {E.}~\bibnamefont
  {Bugaev}}\ and\ \bibinfo {author} {\bibfnamefont {P.}~\bibnamefont
  {Klimai}},\ }\href {\doibase 10.1103/PhysRevD.81.023517} {\bibfield
  {journal} {\bibinfo  {journal} {Phys. Rev.}\ }\textbf {\bibinfo {volume}
  {D81}},\ \bibinfo {pages} {023517} (\bibinfo {year} {2010})},\ \Eprint
  {http://arxiv.org/abs/0908.0664} {arXiv:0908.0664 [astro-ph.CO]} \BibitemShut
  {NoStop}%
%%CITATION = ARXIV:0908.0664;%%
\bibitem [{\citenamefont {Alabidi}\ \emph {et~al.}(2012)\citenamefont
  {Alabidi}, \citenamefont {Kohri}, \citenamefont {Sasaki},\ and\ \citenamefont
  {Sendouda}}]{Alabidi:2012ex}%
  \BibitemOpen
  \bibfield  {author} {\bibinfo {author} {\bibfnamefont {L.}~\bibnamefont
  {Alabidi}}, \bibinfo {author} {\bibfnamefont {K.}~\bibnamefont {Kohri}},
  \bibinfo {author} {\bibfnamefont {M.}~\bibnamefont {Sasaki}}, \ and\ \bibinfo
  {author} {\bibfnamefont {Y.}~\bibnamefont {Sendouda}},\ }\href {\doibase
  10.1088/1475-7516/2012/09/017} {\bibfield  {journal} {\bibinfo  {journal}
  {JCAP}\ }\textbf {\bibinfo {volume} {1209}},\ \bibinfo {pages} {017}
  (\bibinfo {year} {2012})},\ \Eprint {http://arxiv.org/abs/1203.4663}
  {arXiv:1203.4663 [astro-ph.CO]} \BibitemShut {NoStop}%
%%CITATION = ARXIV:1203.4663;%%
\bibitem [{\citenamefont {Maggiore}(2000)}]{Maggiore:1999vm}%
  \BibitemOpen
  \bibfield  {author} {\bibinfo {author} {\bibfnamefont {M.}~\bibnamefont
  {Maggiore}},\ }\href {\doibase 10.1016/S0370-1573(99)00102-7} {\bibfield
  {journal} {\bibinfo  {journal} {Phys. Rept.}\ }\textbf {\bibinfo {volume}
  {331}},\ \bibinfo {pages} {283} (\bibinfo {year} {2000})},\ \Eprint
  {http://arxiv.org/abs/gr-qc/9909001} {arXiv:gr-qc/9909001 [gr-qc]}
  \BibitemShut {NoStop}%
%%CITATION = GR-QC/9909001;%%
\bibitem [{\citenamefont {Maggiore}(2007)}]{Maggiore:1900zz}%
  \BibitemOpen
  \bibfield  {author} {\bibinfo {author} {\bibfnamefont {M.}~\bibnamefont
  {Maggiore}},\ }\href {http://www.oup.com/uk/catalogue/?ci=9780198570745}
  {\emph {\bibinfo {title} {{Gravitational Waves. Vol. 1: Theory and
  Experiments}}}},\ Oxford Master Series in Physics\ (\bibinfo  {publisher}
  {Oxford University Press},\ \bibinfo {year} {2007})\BibitemShut {NoStop}%
%%CITATION = INSPIRE-768483;%%
\bibitem [{\citenamefont {Gundlach}(1999)}]{Gundlach:1999cu}%
  \BibitemOpen
  \bibfield  {author} {\bibinfo {author} {\bibfnamefont {C.}~\bibnamefont
  {Gundlach}},\ }\href@noop {} {\bibfield  {journal} {\bibinfo  {journal}
  {Living Rev. Rel.}\ }\textbf {\bibinfo {volume} {2}},\ \bibinfo {pages} {4}
  (\bibinfo {year} {1999})},\ \Eprint {http://arxiv.org/abs/gr-qc/0001046}
  {arXiv:gr-qc/0001046 [gr-qc]} \BibitemShut {NoStop}%
%%CITATION = GR-QC/0001046;%%
\bibitem [{\citenamefont {Gundlach}(2003)}]{Gundlach:2002sx}%
  \BibitemOpen
  \bibfield  {author} {\bibinfo {author} {\bibfnamefont {C.}~\bibnamefont
  {Gundlach}},\ }\href {\doibase 10.1016/S0370-1573(02)00560-4} {\bibfield
  {journal} {\bibinfo  {journal} {Phys. Rept.}\ }\textbf {\bibinfo {volume}
  {376}},\ \bibinfo {pages} {339} (\bibinfo {year} {2003})},\ \Eprint
  {http://arxiv.org/abs/gr-qc/0210101} {arXiv:gr-qc/0210101 [gr-qc]}
  \BibitemShut {NoStop}%
%%CITATION = GR-QC/0210101;%%
\bibitem [{\citenamefont {Niemeyer}\ and\ \citenamefont
  {Jedamzik}(1998)}]{Niemeyer:1997mt}%
  \BibitemOpen
  \bibfield  {author} {\bibinfo {author} {\bibfnamefont {J.~C.}\ \bibnamefont
  {Niemeyer}}\ and\ \bibinfo {author} {\bibfnamefont {K.}~\bibnamefont
  {Jedamzik}},\ }\href {\doibase 10.1103/PhysRevLett.80.5481} {\bibfield
  {journal} {\bibinfo  {journal} {Phys. Rev. Lett.}\ }\textbf {\bibinfo
  {volume} {80}},\ \bibinfo {pages} {5481} (\bibinfo {year} {1998})},\ \Eprint
  {http://arxiv.org/abs/astro-ph/9709072} {arXiv:astro-ph/9709072 [astro-ph]}
  \BibitemShut {NoStop}%
%%CITATION = ASTRO-PH/9709072;%%
\bibitem [{\citenamefont {Niemeyer}\ and\ \citenamefont
  {Jedamzik}(1999)}]{Niemeyer:1999ak}%
  \BibitemOpen
  \bibfield  {author} {\bibinfo {author} {\bibfnamefont {J.~C.}\ \bibnamefont
  {Niemeyer}}\ and\ \bibinfo {author} {\bibfnamefont {K.}~\bibnamefont
  {Jedamzik}},\ }\href {\doibase 10.1103/PhysRevD.59.124013} {\bibfield
  {journal} {\bibinfo  {journal} {Phys. Rev.}\ }\textbf {\bibinfo {volume}
  {D59}},\ \bibinfo {pages} {124013} (\bibinfo {year} {1999})},\ \Eprint
  {http://arxiv.org/abs/astro-ph/9901292} {arXiv:astro-ph/9901292 [astro-ph]}
  \BibitemShut {NoStop}%
%%CITATION = ASTRO-PH/9901292;%%
\bibitem [{\citenamefont {Kühnel}\ \emph {et~al.}(2016)\citenamefont
  {Kühnel}, \citenamefont {Rampf},\ and\ \citenamefont
  {Sandstad}}]{Kuhnel:2015vtw}%
  \BibitemOpen
  \bibfield  {author} {\bibinfo {author} {\bibfnamefont {F.}~\bibnamefont
  {Kühnel}}, \bibinfo {author} {\bibfnamefont {C.}~\bibnamefont {Rampf}}, \
  and\ \bibinfo {author} {\bibfnamefont {M.}~\bibnamefont {Sandstad}},\ }\href
  {\doibase 10.1140/epjc/s10052-016-3945-8} {\bibfield  {journal} {\bibinfo
  {journal} {Eur. Phys. J.}\ }\textbf {\bibinfo {volume} {C76}},\ \bibinfo
  {pages} {93} (\bibinfo {year} {2016})},\ \Eprint
  {http://arxiv.org/abs/1512.00488} {arXiv:1512.00488 [astro-ph.CO]}
  \BibitemShut {NoStop}%
%%CITATION = ARXIV:1512.00488;%%
\bibitem [{\citenamefont {Dolgov}\ and\ \citenamefont
  {Silk}(1993)}]{Dolgov:1992pu}%
  \BibitemOpen
  \bibfield  {author} {\bibinfo {author} {\bibfnamefont {A.}~\bibnamefont
  {Dolgov}}\ and\ \bibinfo {author} {\bibfnamefont {J.}~\bibnamefont {Silk}},\
  }\href {\doibase 10.1103/PhysRevD.47.4244} {\bibfield  {journal} {\bibinfo
  {journal} {Phys. Rev.}\ }\textbf {\bibinfo {volume} {D47}},\ \bibinfo {pages}
  {4244} (\bibinfo {year} {1993})}\BibitemShut {NoStop}%
%%CITATION = PHRVA,D47,4244;%%
\bibitem [{\citenamefont {Garcia-Bellido}\ \emph {et~al.}(1996)\citenamefont
  {Garcia-Bellido}, \citenamefont {Linde},\ and\ \citenamefont
  {Wands}}]{GarciaBellido:1996qt}%
  \BibitemOpen
  \bibfield  {author} {\bibinfo {author} {\bibfnamefont {J.}~\bibnamefont
  {Garcia-Bellido}}, \bibinfo {author} {\bibfnamefont {A.~D.}\ \bibnamefont
  {Linde}}, \ and\ \bibinfo {author} {\bibfnamefont {D.}~\bibnamefont
  {Wands}},\ }\href {\doibase 10.1103/PhysRevD.54.6040} {\bibfield  {journal}
  {\bibinfo  {journal} {Phys. Rev.}\ }\textbf {\bibinfo {volume} {D54}},\
  \bibinfo {pages} {6040} (\bibinfo {year} {1996})},\ \Eprint
  {http://arxiv.org/abs/astro-ph/9605094} {arXiv:astro-ph/9605094 [astro-ph]}
  \BibitemShut {NoStop}%
%%CITATION = ASTRO-PH/9605094;%%
\bibitem [{\citenamefont {Clesse}\ and\ \citenamefont
  {García-Bellido}(2015)}]{Clesse:2015wea}%
  \BibitemOpen
  \bibfield  {author} {\bibinfo {author} {\bibfnamefont {S.}~\bibnamefont
  {Clesse}}\ and\ \bibinfo {author} {\bibfnamefont {J.}~\bibnamefont
  {García-Bellido}},\ }\href {\doibase 10.1103/PhysRevD.92.023524} {\bibfield
  {journal} {\bibinfo  {journal} {Phys. Rev.}\ }\textbf {\bibinfo {volume}
  {D92}},\ \bibinfo {pages} {023524} (\bibinfo {year} {2015})},\ \Eprint
  {http://arxiv.org/abs/1501.07565} {arXiv:1501.07565 [astro-ph.CO]}
  \BibitemShut {NoStop}%
%%CITATION = ARXIV:1501.07565;%%
\bibitem [{\citenamefont {Cheng}\ \emph {et~al.}(2017)\citenamefont {Cheng},
  \citenamefont {Lee},\ and\ \citenamefont {Ng}}]{Cheng:2016qzb}%
  \BibitemOpen
  \bibfield  {author} {\bibinfo {author} {\bibfnamefont {S.-L.}\ \bibnamefont
  {Cheng}}, \bibinfo {author} {\bibfnamefont {W.}~\bibnamefont {Lee}}, \ and\
  \bibinfo {author} {\bibfnamefont {K.-W.}\ \bibnamefont {Ng}},\ }\href
  {\doibase 10.1007/JHEP02(2017)008} {\bibfield  {journal} {\bibinfo  {journal}
  {JHEP}\ }\textbf {\bibinfo {volume} {02}},\ \bibinfo {pages} {008} (\bibinfo
  {year} {2017})},\ \Eprint {http://arxiv.org/abs/1606.00206} {arXiv:1606.00206
  [astro-ph.CO]} \BibitemShut {NoStop}%
%%CITATION = ARXIV:1606.00206;%%
\bibitem [{\citenamefont {Blinnikov}\ \emph {et~al.}(2016)\citenamefont
  {Blinnikov}, \citenamefont {Dolgov}, \citenamefont {Porayko},\ and\
  \citenamefont {Postnov}}]{Blinnikov:2016bxu}%
  \BibitemOpen
  \bibfield  {author} {\bibinfo {author} {\bibfnamefont {S.}~\bibnamefont
  {Blinnikov}}, \bibinfo {author} {\bibfnamefont {A.}~\bibnamefont {Dolgov}},
  \bibinfo {author} {\bibfnamefont {N.~K.}\ \bibnamefont {Porayko}}, \ and\
  \bibinfo {author} {\bibfnamefont {K.}~\bibnamefont {Postnov}},\ }\href
  {\doibase 10.1088/1475-7516/2016/11/036} {\bibfield  {journal} {\bibinfo
  {journal} {JCAP}\ }\textbf {\bibinfo {volume} {1611}},\ \bibinfo {pages}
  {036} (\bibinfo {year} {2016})},\ \Eprint {http://arxiv.org/abs/1611.00541}
  {arXiv:1611.00541 [astro-ph.HE]} \BibitemShut {NoStop}%
%%CITATION = ARXIV:1611.00541;%%
\bibitem [{\citenamefont {Lentati}\ \emph {et~al.}(2015)\citenamefont {Lentati}
  \emph {et~al.}}]{Lentati:2015qwp}%
  \BibitemOpen
  \bibfield  {author} {\bibinfo {author} {\bibfnamefont {L.}~\bibnamefont
  {Lentati}} \emph {et~al.},\ }\href {\doibase 10.1093/mnras/stv1538}
  {\bibfield  {journal} {\bibinfo  {journal} {Mon. Not. Roy. Astron. Soc.}\
  }\textbf {\bibinfo {volume} {453}},\ \bibinfo {pages} {2576} (\bibinfo {year}
  {2015})},\ \Eprint {http://arxiv.org/abs/1504.03692} {arXiv:1504.03692
  [astro-ph.CO]} \BibitemShut {NoStop}%
%%CITATION = ARXIV:1504.03692;%%
\bibitem [{\citenamefont {Shannon}\ \emph {et~al.}(2015)\citenamefont {Shannon}
  \emph {et~al.}}]{Shannon:2015ect}%
  \BibitemOpen
  \bibfield  {author} {\bibinfo {author} {\bibfnamefont {R.~M.}\ \bibnamefont
  {Shannon}} \emph {et~al.},\ }\href {\doibase 10.1126/science.aab1910}
  {\bibfield  {journal} {\bibinfo  {journal} {Science}\ }\textbf {\bibinfo
  {volume} {349}},\ \bibinfo {pages} {1522} (\bibinfo {year} {2015})},\ \Eprint
  {http://arxiv.org/abs/1509.07320} {arXiv:1509.07320 [astro-ph.CO]}
  \BibitemShut {NoStop}%
%%CITATION = ARXIV:1509.07320;%%
\bibitem [{\citenamefont {Arzoumanian}\ \emph {et~al.}(2016)\citenamefont
  {Arzoumanian} \emph {et~al.}}]{Arzoumanian:2015liz}%
  \BibitemOpen
  \bibfield  {author} {\bibinfo {author} {\bibfnamefont {Z.}~\bibnamefont
  {Arzoumanian}} \emph {et~al.} (\bibinfo {collaboration} {NANOGrav}),\ }\href
  {\doibase 10.3847/0004-637X/821/1/13} {\bibfield  {journal} {\bibinfo
  {journal} {Astrophys. J.}\ }\textbf {\bibinfo {volume} {821}},\ \bibinfo
  {pages} {13} (\bibinfo {year} {2016})},\ \Eprint
  {http://arxiv.org/abs/1508.03024} {arXiv:1508.03024 [astro-ph.GA]}
  \BibitemShut {NoStop}%
%%CITATION = ARXIV:1508.03024;%%
\bibitem [{\citenamefont {Janssen}\ \emph {et~al.}(2015)\citenamefont {Janssen}
  \emph {et~al.}}]{Janssen:2014dka}%
  \BibitemOpen
  \bibfield  {author} {\bibinfo {author} {\bibfnamefont {G.}~\bibnamefont
  {Janssen}} \emph {et~al.},\ }\bibfield  {booktitle} {\emph {\bibinfo
  {booktitle} {{Proceedings, Advancing Astrophysics with the Square Kilometre
  Array (AASKA14): Giardini Naxos, Italy, June 9-13, 2014}}},\ }\href@noop {}
  {\bibfield  {journal} {\bibinfo  {journal} {PoS}\ }\textbf {\bibinfo {volume}
  {AASKA14}},\ \bibinfo {pages} {037} (\bibinfo {year} {2015})},\ \Eprint
  {http://arxiv.org/abs/1501.00127} {arXiv:1501.00127 [astro-ph.IM]}
  \BibitemShut {NoStop}%
%%CITATION = ARXIV:1501.00127;%%
\bibitem [{\citenamefont {Kofman}\ \emph {et~al.}(1994)\citenamefont {Kofman},
  \citenamefont {Linde},\ and\ \citenamefont {Starobinsky}}]{Kofman:1994rk}%
  \BibitemOpen
  \bibfield  {author} {\bibinfo {author} {\bibfnamefont {L.}~\bibnamefont
  {Kofman}}, \bibinfo {author} {\bibfnamefont {A.~D.}\ \bibnamefont {Linde}}, \
  and\ \bibinfo {author} {\bibfnamefont {A.~A.}\ \bibnamefont {Starobinsky}},\
  }\href {\doibase 10.1103/PhysRevLett.73.3195} {\bibfield  {journal} {\bibinfo
   {journal} {Phys. Rev. Lett.}\ }\textbf {\bibinfo {volume} {73}},\ \bibinfo
  {pages} {3195} (\bibinfo {year} {1994})},\ \Eprint
  {http://arxiv.org/abs/hep-th/9405187} {arXiv:hep-th/9405187 [hep-th]}
  \BibitemShut {NoStop}%
%%CITATION = HEP-TH/9405187;%%
\bibitem [{\citenamefont {Kawasaki}\ \emph {et~al.}(2006)\citenamefont
  {Kawasaki}, \citenamefont {Takayama}, \citenamefont {Yamaguchi},\ and\
  \citenamefont {Yokoyama}}]{Kawasaki:2006zv}%
  \BibitemOpen
  \bibfield  {author} {\bibinfo {author} {\bibfnamefont {M.}~\bibnamefont
  {Kawasaki}}, \bibinfo {author} {\bibfnamefont {T.}~\bibnamefont {Takayama}},
  \bibinfo {author} {\bibfnamefont {M.}~\bibnamefont {Yamaguchi}}, \ and\
  \bibinfo {author} {\bibfnamefont {J.}~\bibnamefont {Yokoyama}},\ }\href
  {\doibase 10.1103/PhysRevD.74.043525} {\bibfield  {journal} {\bibinfo
  {journal} {Phys. Rev.}\ }\textbf {\bibinfo {volume} {D74}},\ \bibinfo {pages}
  {043525} (\bibinfo {year} {2006})},\ \Eprint
  {http://arxiv.org/abs/hep-ph/0605271} {arXiv:hep-ph/0605271 [hep-ph]}
  \BibitemShut {NoStop}%
%%CITATION = HEP-PH/0605271;%%
\bibitem [{\citenamefont {Kawaguchi}\ \emph {et~al.}(2008)\citenamefont
  {Kawaguchi}, \citenamefont {Kawasaki}, \citenamefont {Takayama},
  \citenamefont {Yamaguchi},\ and\ \citenamefont
  {Yokoyama}}]{Kawaguchi:2007fz}%
  \BibitemOpen
  \bibfield  {author} {\bibinfo {author} {\bibfnamefont {T.}~\bibnamefont
  {Kawaguchi}}, \bibinfo {author} {\bibfnamefont {M.}~\bibnamefont {Kawasaki}},
  \bibinfo {author} {\bibfnamefont {T.}~\bibnamefont {Takayama}}, \bibinfo
  {author} {\bibfnamefont {M.}~\bibnamefont {Yamaguchi}}, \ and\ \bibinfo
  {author} {\bibfnamefont {J.}~\bibnamefont {Yokoyama}},\ }\href {\doibase
  10.1111/j.1365-2966.2008.13523.x} {\bibfield  {journal} {\bibinfo  {journal}
  {Mon. Not. Roy. Astron. Soc.}\ }\textbf {\bibinfo {volume} {388}},\ \bibinfo
  {pages} {1426} (\bibinfo {year} {2008})},\ \Eprint
  {http://arxiv.org/abs/0711.3886} {arXiv:0711.3886 [astro-ph]} \BibitemShut
  {NoStop}%
%%CITATION = ARXIV:0711.3886;%%
\bibitem [{\citenamefont {Frampton}\ \emph {et~al.}(2010)\citenamefont
  {Frampton}, \citenamefont {Kawasaki}, \citenamefont {Takahashi},\ and\
  \citenamefont {Yanagida}}]{Frampton:2010sw}%
  \BibitemOpen
  \bibfield  {author} {\bibinfo {author} {\bibfnamefont {P.~H.}\ \bibnamefont
  {Frampton}}, \bibinfo {author} {\bibfnamefont {M.}~\bibnamefont {Kawasaki}},
  \bibinfo {author} {\bibfnamefont {F.}~\bibnamefont {Takahashi}}, \ and\
  \bibinfo {author} {\bibfnamefont {T.~T.}\ \bibnamefont {Yanagida}},\ }\href
  {\doibase 10.1088/1475-7516/2010/04/023} {\bibfield  {journal} {\bibinfo
  {journal} {JCAP}\ }\textbf {\bibinfo {volume} {1004}},\ \bibinfo {pages}
  {023} (\bibinfo {year} {2010})},\ \Eprint {http://arxiv.org/abs/1001.2308}
  {arXiv:1001.2308 [hep-ph]} \BibitemShut {NoStop}%
%%CITATION = ARXIV:1001.2308;%%
\bibitem [{\citenamefont {Copeland}\ \emph {et~al.}(1994)\citenamefont
  {Copeland}, \citenamefont {Liddle}, \citenamefont {Lyth}, \citenamefont
  {Stewart},\ and\ \citenamefont {Wands}}]{Copeland:1994vg}%
  \BibitemOpen
  \bibfield  {author} {\bibinfo {author} {\bibfnamefont {E.~J.}\ \bibnamefont
  {Copeland}}, \bibinfo {author} {\bibfnamefont {A.~R.}\ \bibnamefont
  {Liddle}}, \bibinfo {author} {\bibfnamefont {D.~H.}\ \bibnamefont {Lyth}},
  \bibinfo {author} {\bibfnamefont {E.~D.}\ \bibnamefont {Stewart}}, \ and\
  \bibinfo {author} {\bibfnamefont {D.}~\bibnamefont {Wands}},\ }\href
  {\doibase 10.1103/PhysRevD.49.6410} {\bibfield  {journal} {\bibinfo
  {journal} {Phys. Rev.}\ }\textbf {\bibinfo {volume} {D49}},\ \bibinfo {pages}
  {6410} (\bibinfo {year} {1994})},\ \Eprint
  {http://arxiv.org/abs/astro-ph/9401011} {arXiv:astro-ph/9401011 [astro-ph]}
  \BibitemShut {NoStop}%
%%CITATION = ASTRO-PH/9401011;%%
\bibitem [{\citenamefont {Linde}\ and\ \citenamefont
  {Riotto}(1997)}]{Linde:1997sj}%
  \BibitemOpen
  \bibfield  {author} {\bibinfo {author} {\bibfnamefont {A.~D.}\ \bibnamefont
  {Linde}}\ and\ \bibinfo {author} {\bibfnamefont {A.}~\bibnamefont {Riotto}},\
  }\href {\doibase 10.1103/PhysRevD.56.R1841} {\bibfield  {journal} {\bibinfo
  {journal} {Phys. Rev.}\ }\textbf {\bibinfo {volume} {D56}},\ \bibinfo {pages}
  {R1841} (\bibinfo {year} {1997})},\ \Eprint
  {http://arxiv.org/abs/hep-ph/9703209} {arXiv:hep-ph/9703209 [hep-ph]}
  \BibitemShut {NoStop}%
%%CITATION = HEP-PH/9703209;%%
\bibitem [{\citenamefont {Izawa}\ and\ \citenamefont
  {Yanagida}(1997)}]{Izawa:1996dv}%
  \BibitemOpen
  \bibfield  {author} {\bibinfo {author} {\bibfnamefont {K.~I.}\ \bibnamefont
  {Izawa}}\ and\ \bibinfo {author} {\bibfnamefont {T.}~\bibnamefont
  {Yanagida}},\ }\href {\doibase 10.1016/S0370-2693(96)01638-3} {\bibfield
  {journal} {\bibinfo  {journal} {Phys. Lett.}\ }\textbf {\bibinfo {volume}
  {B393}},\ \bibinfo {pages} {331} (\bibinfo {year} {1997})},\ \Eprint
  {http://arxiv.org/abs/hep-ph/9608359} {arXiv:hep-ph/9608359 [hep-ph]}
  \BibitemShut {NoStop}%
%%CITATION = HEP-PH/9608359;%%
\bibitem [{\citenamefont {Izawa}\ \emph {et~al.}(1997)\citenamefont {Izawa},
  \citenamefont {Kawasaki},\ and\ \citenamefont {Yanagida}}]{Izawa:1997df}%
  \BibitemOpen
  \bibfield  {author} {\bibinfo {author} {\bibfnamefont {K.~I.}\ \bibnamefont
  {Izawa}}, \bibinfo {author} {\bibfnamefont {M.}~\bibnamefont {Kawasaki}}, \
  and\ \bibinfo {author} {\bibfnamefont {T.}~\bibnamefont {Yanagida}},\ }\href
  {\doibase 10.1016/S0370-2693(97)01040-X} {\bibfield  {journal} {\bibinfo
  {journal} {Phys. Lett.}\ }\textbf {\bibinfo {volume} {B411}},\ \bibinfo
  {pages} {249} (\bibinfo {year} {1997})},\ \Eprint
  {http://arxiv.org/abs/hep-ph/9707201} {arXiv:hep-ph/9707201 [hep-ph]}
  \BibitemShut {NoStop}%
%%CITATION = HEP-PH/9707201;%%
\bibitem [{\citenamefont {Kofman}\ \emph {et~al.}(1997)\citenamefont {Kofman},
  \citenamefont {Linde},\ and\ \citenamefont {Starobinsky}}]{Kofman:1997yn}%
  \BibitemOpen
  \bibfield  {author} {\bibinfo {author} {\bibfnamefont {L.}~\bibnamefont
  {Kofman}}, \bibinfo {author} {\bibfnamefont {A.~D.}\ \bibnamefont {Linde}}, \
  and\ \bibinfo {author} {\bibfnamefont {A.~A.}\ \bibnamefont {Starobinsky}},\
  }\href {\doibase 10.1103/PhysRevD.56.3258} {\bibfield  {journal} {\bibinfo
  {journal} {Phys. Rev.}\ }\textbf {\bibinfo {volume} {D56}},\ \bibinfo {pages}
  {3258} (\bibinfo {year} {1997})},\ \Eprint
  {http://arxiv.org/abs/hep-ph/9704452} {arXiv:hep-ph/9704452 [hep-ph]}
  \BibitemShut {NoStop}%
%%CITATION = HEP-PH/9704452;%%
\bibitem [{\citenamefont {Kawasaki}\ \emph {et~al.}(1998)\citenamefont
  {Kawasaki}, \citenamefont {Sugiyama},\ and\ \citenamefont
  {Yanagida}}]{Kawasaki:1997ju}%
  \BibitemOpen
  \bibfield  {author} {\bibinfo {author} {\bibfnamefont {M.}~\bibnamefont
  {Kawasaki}}, \bibinfo {author} {\bibfnamefont {N.}~\bibnamefont {Sugiyama}},
  \ and\ \bibinfo {author} {\bibfnamefont {T.}~\bibnamefont {Yanagida}},\
  }\href {\doibase 10.1103/PhysRevD.57.6050} {\bibfield  {journal} {\bibinfo
  {journal} {Phys. Rev.}\ }\textbf {\bibinfo {volume} {D57}},\ \bibinfo {pages}
  {6050} (\bibinfo {year} {1998})},\ \Eprint
  {http://arxiv.org/abs/hep-ph/9710259} {arXiv:hep-ph/9710259 [hep-ph]}
  \BibitemShut {NoStop}%
%%CITATION = HEP-PH/9710259;%%
\bibitem [{\citenamefont {Kawasaki}\ and\ \citenamefont
  {Yanagida}(1999)}]{Kawasaki:1998vx}%
  \BibitemOpen
  \bibfield  {author} {\bibinfo {author} {\bibfnamefont {M.}~\bibnamefont
  {Kawasaki}}\ and\ \bibinfo {author} {\bibfnamefont {T.}~\bibnamefont
  {Yanagida}},\ }\href {\doibase 10.1103/PhysRevD.59.043512} {\bibfield
  {journal} {\bibinfo  {journal} {Phys. Rev.}\ }\textbf {\bibinfo {volume}
  {D59}},\ \bibinfo {pages} {043512} (\bibinfo {year} {1999})},\ \Eprint
  {http://arxiv.org/abs/hep-ph/9807544} {arXiv:hep-ph/9807544 [hep-ph]}
  \BibitemShut {NoStop}%
%%CITATION = HEP-PH/9807544;%%
\bibitem [{\citenamefont {Stewart}(1997{\natexlab{a}})}]{Stewart:1996ey}%
  \BibitemOpen
  \bibfield  {author} {\bibinfo {author} {\bibfnamefont {E.~D.}\ \bibnamefont
  {Stewart}},\ }\href {\doibase 10.1016/S0370-2693(96)01458-X} {\bibfield
  {journal} {\bibinfo  {journal} {Phys. Lett.}\ }\textbf {\bibinfo {volume}
  {B391}},\ \bibinfo {pages} {34} (\bibinfo {year} {1997}{\natexlab{a}})},\
  \Eprint {http://arxiv.org/abs/hep-ph/9606241} {arXiv:hep-ph/9606241 [hep-ph]}
  \BibitemShut {NoStop}%
%%CITATION = HEP-PH/9606241;%%
\bibitem [{\citenamefont {Stewart}(1997{\natexlab{b}})}]{Stewart:1997wg}%
  \BibitemOpen
  \bibfield  {author} {\bibinfo {author} {\bibfnamefont {E.~D.}\ \bibnamefont
  {Stewart}},\ }\href {\doibase 10.1103/PhysRevD.56.2019} {\bibfield  {journal}
  {\bibinfo  {journal} {Phys. Rev.}\ }\textbf {\bibinfo {volume} {D56}},\
  \bibinfo {pages} {2019} (\bibinfo {year} {1997}{\natexlab{b}})},\ \Eprint
  {http://arxiv.org/abs/hep-ph/9703232} {arXiv:hep-ph/9703232 [hep-ph]}
  \BibitemShut {NoStop}%
%%CITATION = HEP-PH/9703232;%%
\bibitem [{\citenamefont {Leach}\ \emph {et~al.}(2000)\citenamefont {Leach},
  \citenamefont {Grivell},\ and\ \citenamefont {Liddle}}]{Leach:2000ea}%
  \BibitemOpen
  \bibfield  {author} {\bibinfo {author} {\bibfnamefont {S.~M.}\ \bibnamefont
  {Leach}}, \bibinfo {author} {\bibfnamefont {I.~J.}\ \bibnamefont {Grivell}},
  \ and\ \bibinfo {author} {\bibfnamefont {A.~R.}\ \bibnamefont {Liddle}},\
  }\href {\doibase 10.1103/PhysRevD.62.043516} {\bibfield  {journal} {\bibinfo
  {journal} {Phys. Rev.}\ }\textbf {\bibinfo {volume} {D62}},\ \bibinfo {pages}
  {043516} (\bibinfo {year} {2000})},\ \Eprint
  {http://arxiv.org/abs/astro-ph/0004296} {arXiv:astro-ph/0004296 [astro-ph]}
  \BibitemShut {NoStop}%
%%CITATION = ASTRO-PH/0004296;%%
\bibitem [{\citenamefont {Drees}\ and\ \citenamefont
  {Erfani}(2011)}]{Drees:2011hb}%
  \BibitemOpen
  \bibfield  {author} {\bibinfo {author} {\bibfnamefont {M.}~\bibnamefont
  {Drees}}\ and\ \bibinfo {author} {\bibfnamefont {E.}~\bibnamefont {Erfani}},\
  }\href {\doibase 10.1088/1475-7516/2011/04/005} {\bibfield  {journal}
  {\bibinfo  {journal} {JCAP}\ }\textbf {\bibinfo {volume} {1104}},\ \bibinfo
  {pages} {005} (\bibinfo {year} {2011})},\ \Eprint
  {http://arxiv.org/abs/1102.2340} {arXiv:1102.2340 [hep-ph]} \BibitemShut
  {NoStop}%
%%CITATION = ARXIV:1102.2340;%%
\bibitem [{\citenamefont {Drees}\ and\ \citenamefont
  {Erfani}(2012)}]{Drees:2011yz}%
  \BibitemOpen
  \bibfield  {author} {\bibinfo {author} {\bibfnamefont {M.}~\bibnamefont
  {Drees}}\ and\ \bibinfo {author} {\bibfnamefont {E.}~\bibnamefont {Erfani}},\
  }\href {\doibase 10.1088/1475-7516/2012/01/035} {\bibfield  {journal}
  {\bibinfo  {journal} {JCAP}\ }\textbf {\bibinfo {volume} {1201}},\ \bibinfo
  {pages} {035} (\bibinfo {year} {2012})},\ \Eprint
  {http://arxiv.org/abs/1110.6052} {arXiv:1110.6052 [astro-ph.CO]} \BibitemShut
  {NoStop}%
%%CITATION = ARXIV:1110.6052;%%
\bibitem [{\citenamefont {Ade}\ \emph {et~al.}(2016{\natexlab{a}})\citenamefont
  {Ade} \emph {et~al.}}]{Ade:2015xua}%
  \BibitemOpen
  \bibfield  {author} {\bibinfo {author} {\bibfnamefont {P.~A.~R.}\
  \bibnamefont {Ade}} \emph {et~al.} (\bibinfo {collaboration} {Planck}),\
  }\href {\doibase 10.1051/0004-6361/201525830} {\bibfield  {journal} {\bibinfo
   {journal} {Astron. Astrophys.}\ }\textbf {\bibinfo {volume} {594}},\
  \bibinfo {pages} {A13} (\bibinfo {year} {2016}{\natexlab{a}})},\ \Eprint
  {http://arxiv.org/abs/1502.01589} {arXiv:1502.01589 [astro-ph.CO]}
  \BibitemShut {NoStop}%
%%CITATION = ARXIV:1502.01589;%%
\bibitem [{\citenamefont {Ade}\ \emph {et~al.}(2016{\natexlab{b}})\citenamefont
  {Ade} \emph {et~al.}}]{Ade:2015lrj}%
  \BibitemOpen
  \bibfield  {author} {\bibinfo {author} {\bibfnamefont {P.~A.~R.}\
  \bibnamefont {Ade}} \emph {et~al.} (\bibinfo {collaboration} {Planck}),\
  }\href {\doibase 10.1051/0004-6361/201525898} {\bibfield  {journal} {\bibinfo
   {journal} {Astron. Astrophys.}\ }\textbf {\bibinfo {volume} {594}},\
  \bibinfo {pages} {A20} (\bibinfo {year} {2016}{\natexlab{b}})},\ \Eprint
  {http://arxiv.org/abs/1502.02114} {arXiv:1502.02114 [astro-ph.CO]}
  \BibitemShut {NoStop}%
%%CITATION = ARXIV:1502.02114;%%
\bibitem [{\citenamefont {Kasuya}\ and\ \citenamefont
  {Kawasaki}(2009)}]{Kasuya:2009up}%
  \BibitemOpen
  \bibfield  {author} {\bibinfo {author} {\bibfnamefont {S.}~\bibnamefont
  {Kasuya}}\ and\ \bibinfo {author} {\bibfnamefont {M.}~\bibnamefont
  {Kawasaki}},\ }\href {\doibase 10.1103/PhysRevD.80.023516} {\bibfield
  {journal} {\bibinfo  {journal} {Phys. Rev.}\ }\textbf {\bibinfo {volume}
  {D80}},\ \bibinfo {pages} {023516} (\bibinfo {year} {2009})},\ \Eprint
  {http://arxiv.org/abs/0904.3800} {arXiv:0904.3800 [astro-ph.CO]} \BibitemShut
  {NoStop}%
%%CITATION = ARXIV:0904.3800;%%
\bibitem [{\citenamefont {Kawasaki}\ \emph {et~al.}(2013)\citenamefont
  {Kawasaki}, \citenamefont {Kitajima},\ and\ \citenamefont
  {Yanagida}}]{Kawasaki:2012wr}%
  \BibitemOpen
  \bibfield  {author} {\bibinfo {author} {\bibfnamefont {M.}~\bibnamefont
  {Kawasaki}}, \bibinfo {author} {\bibfnamefont {N.}~\bibnamefont {Kitajima}},
  \ and\ \bibinfo {author} {\bibfnamefont {T.~T.}\ \bibnamefont {Yanagida}},\
  }\href {\doibase 10.1103/PhysRevD.87.063519} {\bibfield  {journal} {\bibinfo
  {journal} {Phys. Rev.}\ }\textbf {\bibinfo {volume} {D87}},\ \bibinfo {pages}
  {063519} (\bibinfo {year} {2013})},\ \Eprint {http://arxiv.org/abs/1207.2550}
  {arXiv:1207.2550 [hep-ph]} \BibitemShut {NoStop}%
%%CITATION = ARXIV:1207.2550;%%
\bibitem [{\citenamefont {Kohri}\ \emph {et~al.}(2013)\citenamefont {Kohri},
  \citenamefont {Lin},\ and\ \citenamefont {Matsuda}}]{Kohri:2012yw}%
  \BibitemOpen
  \bibfield  {author} {\bibinfo {author} {\bibfnamefont {K.}~\bibnamefont
  {Kohri}}, \bibinfo {author} {\bibfnamefont {C.-M.}\ \bibnamefont {Lin}}, \
  and\ \bibinfo {author} {\bibfnamefont {T.}~\bibnamefont {Matsuda}},\ }\href
  {\doibase 10.1103/PhysRevD.87.103527} {\bibfield  {journal} {\bibinfo
  {journal} {Phys. Rev.}\ }\textbf {\bibinfo {volume} {D87}},\ \bibinfo {pages}
  {103527} (\bibinfo {year} {2013})},\ \Eprint {http://arxiv.org/abs/1211.2371}
  {arXiv:1211.2371 [hep-ph]} \BibitemShut {NoStop}%
%%CITATION = ARXIV:1211.2371;%%
\bibitem [{\citenamefont {Lyth}\ and\ \citenamefont
  {Wands}(2002)}]{Lyth:2001nq}%
  \BibitemOpen
  \bibfield  {author} {\bibinfo {author} {\bibfnamefont {D.~H.}\ \bibnamefont
  {Lyth}}\ and\ \bibinfo {author} {\bibfnamefont {D.}~\bibnamefont {Wands}},\
  }\href {\doibase 10.1016/S0370-2693(01)01366-1} {\bibfield  {journal}
  {\bibinfo  {journal} {Phys. Lett.}\ }\textbf {\bibinfo {volume} {B524}},\
  \bibinfo {pages} {5} (\bibinfo {year} {2002})},\ \Eprint
  {http://arxiv.org/abs/hep-ph/0110002} {arXiv:hep-ph/0110002 [hep-ph]}
  \BibitemShut {NoStop}%
%%CITATION = HEP-PH/0110002;%%
\bibitem [{\citenamefont {Lyth}\ \emph {et~al.}(2003)\citenamefont {Lyth},
  \citenamefont {Ungarelli},\ and\ \citenamefont {Wands}}]{Lyth:2002my}%
  \BibitemOpen
  \bibfield  {author} {\bibinfo {author} {\bibfnamefont {D.~H.}\ \bibnamefont
  {Lyth}}, \bibinfo {author} {\bibfnamefont {C.}~\bibnamefont {Ungarelli}}, \
  and\ \bibinfo {author} {\bibfnamefont {D.}~\bibnamefont {Wands}},\ }\href
  {\doibase 10.1103/PhysRevD.67.023503} {\bibfield  {journal} {\bibinfo
  {journal} {Phys. Rev.}\ }\textbf {\bibinfo {volume} {D67}},\ \bibinfo {pages}
  {023503} (\bibinfo {year} {2003})},\ \Eprint
  {http://arxiv.org/abs/astro-ph/0208055} {arXiv:astro-ph/0208055 [astro-ph]}
  \BibitemShut {NoStop}%
%%CITATION = ASTRO-PH/0208055;%%
\bibitem [{\citenamefont {Khlopov}\ and\ \citenamefont
  {Polnarev}(1980)}]{Khlopov:1980mg}%
  \BibitemOpen
  \bibfield  {author} {\bibinfo {author} {\bibfnamefont {M.~{\relax Yu}.}\
  \bibnamefont {Khlopov}}\ and\ \bibinfo {author} {\bibfnamefont {A.~G.}\
  \bibnamefont {Polnarev}},\ }\href {\doibase 10.1016/0370-2693(80)90624-3}
  {\bibfield  {journal} {\bibinfo  {journal} {Phys. Lett.}\ }\textbf {\bibinfo
  {volume} {B97}},\ \bibinfo {pages} {383} (\bibinfo {year}
  {1980})}\BibitemShut {NoStop}%
%%CITATION = PHLTA,B97,383;%%
\bibitem [{\citenamefont {Jedamzik}(1997)}]{Jedamzik:1996mr}%
  \BibitemOpen
  \bibfield  {author} {\bibinfo {author} {\bibfnamefont {K.}~\bibnamefont
  {Jedamzik}},\ }\href {\doibase 10.1103/PhysRevD.55.5871} {\bibfield
  {journal} {\bibinfo  {journal} {Phys. Rev.}\ }\textbf {\bibinfo {volume}
  {D55}},\ \bibinfo {pages} {5871} (\bibinfo {year} {1997})},\ \Eprint
  {http://arxiv.org/abs/astro-ph/9605152} {arXiv:astro-ph/9605152 [astro-ph]}
  \BibitemShut {NoStop}%
%%CITATION = ASTRO-PH/9605152;%%
\bibitem [{\citenamefont {Jedamzik}(1998)}]{Jedamzik:1998hc}%
  \BibitemOpen
  \bibfield  {author} {\bibinfo {author} {\bibfnamefont {K.}~\bibnamefont
  {Jedamzik}},\ }\bibfield  {booktitle} {\emph {\bibinfo {booktitle} {{Sources
  and detection of dark matter in the universe. Proceedings, 3rd International
  Symposium, and Workshop on Primordial Black Holes and Hawking Radiation,
  Marina del Rey, USA, February 17-20, 1998}}},\ }\href {\doibase
  10.1016/S0370-1573(98)00067-2} {\bibfield  {journal} {\bibinfo  {journal}
  {Phys. Rept.}\ }\textbf {\bibinfo {volume} {307}},\ \bibinfo {pages} {155}
  (\bibinfo {year} {1998})},\ \Eprint {http://arxiv.org/abs/astro-ph/9805147}
  {arXiv:astro-ph/9805147 [astro-ph]} \BibitemShut {NoStop}%
%%CITATION = ASTRO-PH/9805147;%%
\bibitem [{\citenamefont {Jedamzik}\ and\ \citenamefont
  {Niemeyer}(1999)}]{Jedamzik:1999am}%
  \BibitemOpen
  \bibfield  {author} {\bibinfo {author} {\bibfnamefont {K.}~\bibnamefont
  {Jedamzik}}\ and\ \bibinfo {author} {\bibfnamefont {J.~C.}\ \bibnamefont
  {Niemeyer}},\ }\href {\doibase 10.1103/PhysRevD.59.124014} {\bibfield
  {journal} {\bibinfo  {journal} {Phys. Rev.}\ }\textbf {\bibinfo {volume}
  {D59}},\ \bibinfo {pages} {124014} (\bibinfo {year} {1999})},\ \Eprint
  {http://arxiv.org/abs/astro-ph/9901293} {arXiv:astro-ph/9901293 [astro-ph]}
  \BibitemShut {NoStop}%
%%CITATION = ASTRO-PH/9901293;%%
\bibitem [{\citenamefont {Phinney}(2001)}]{Phinney:2001di}%
  \BibitemOpen
  \bibfield  {author} {\bibinfo {author} {\bibfnamefont {E.~S.}\ \bibnamefont
  {Phinney}},\ }\href@noop {} {\  (\bibinfo {year} {2001})},\ \Eprint
  {http://arxiv.org/abs/astro-ph/0108028} {arXiv:astro-ph/0108028 [astro-ph]}
  \BibitemShut {NoStop}%
%%CITATION = ASTRO-PH/0108028;%%
\bibitem [{\citenamefont {Jaffe}\ and\ \citenamefont
  {Backer}(2003)}]{Jaffe:2002rt}%
  \BibitemOpen
  \bibfield  {author} {\bibinfo {author} {\bibfnamefont {A.~H.}\ \bibnamefont
  {Jaffe}}\ and\ \bibinfo {author} {\bibfnamefont {D.~C.}\ \bibnamefont
  {Backer}},\ }\href {\doibase 10.1086/345443} {\bibfield  {journal} {\bibinfo
  {journal} {Astrophys. J.}\ }\textbf {\bibinfo {volume} {583}},\ \bibinfo
  {pages} {616} (\bibinfo {year} {2003})},\ \Eprint
  {http://arxiv.org/abs/astro-ph/0210148} {arXiv:astro-ph/0210148 [astro-ph]}
  \BibitemShut {NoStop}%
%%CITATION = ASTRO-PH/0210148;%%
\bibitem [{\citenamefont {Wyithe}\ and\ \citenamefont
  {Loeb}(2003)}]{Wyithe:2002ep}%
  \BibitemOpen
  \bibfield  {author} {\bibinfo {author} {\bibfnamefont {J.~S.~B.}\
  \bibnamefont {Wyithe}}\ and\ \bibinfo {author} {\bibfnamefont
  {A.}~\bibnamefont {Loeb}},\ }\href {\doibase 10.1086/375187} {\bibfield
  {journal} {\bibinfo  {journal} {Astrophys. J.}\ }\textbf {\bibinfo {volume}
  {590}},\ \bibinfo {pages} {691} (\bibinfo {year} {2003})},\ \Eprint
  {http://arxiv.org/abs/astro-ph/0211556} {arXiv:astro-ph/0211556 [astro-ph]}
  \BibitemShut {NoStop}%
%%CITATION = ASTRO-PH/0211556;%%
\bibitem [{\citenamefont {Clesse}\ and\ \citenamefont
  {García-Bellido}(2016{\natexlab{a}})}]{Clesse:2016vqa}%
  \BibitemOpen
  \bibfield  {author} {\bibinfo {author} {\bibfnamefont {S.}~\bibnamefont
  {Clesse}}\ and\ \bibinfo {author} {\bibfnamefont {J.}~\bibnamefont
  {García-Bellido}},\ }\href {\doibase 10.1016/j.dark.2016.10.002} {\bibfield
  {journal} {\bibinfo  {journal} {Phys. Dark Univ.}\ }\textbf {\bibinfo
  {volume} {10}},\ \bibinfo {pages} {002} (\bibinfo {year}
  {2016}{\natexlab{a}})},\ \Eprint {http://arxiv.org/abs/1603.05234}
  {arXiv:1603.05234 [astro-ph.CO]} \BibitemShut {NoStop}%
%%CITATION = ARXIV:1603.05234;%%
\bibitem [{\citenamefont {Clesse}\ and\ \citenamefont
  {García-Bellido}(2016{\natexlab{b}})}]{Clesse:2016ajp}%
  \BibitemOpen
  \bibfield  {author} {\bibinfo {author} {\bibfnamefont {S.}~\bibnamefont
  {Clesse}}\ and\ \bibinfo {author} {\bibfnamefont {J.}~\bibnamefont
  {García-Bellido}},\ }\href@noop {} {\  (\bibinfo {year}
  {2016}{\natexlab{b}})},\ \Eprint {http://arxiv.org/abs/1610.08479}
  {arXiv:1610.08479 [astro-ph.CO]} \BibitemShut {NoStop}%
%%CITATION = ARXIV:1610.08479;%%
\bibitem [{\citenamefont {Cholis}\ \emph {et~al.}(2016)\citenamefont {Cholis},
  \citenamefont {Kovetz}, \citenamefont {Ali-Haïmoud}, \citenamefont {Bird},
  \citenamefont {Kamionkowski}, \citenamefont {Muñoz},\ and\ \citenamefont
  {Raccanelli}}]{Cholis:2016kqi}%
  \BibitemOpen
  \bibfield  {author} {\bibinfo {author} {\bibfnamefont {I.}~\bibnamefont
  {Cholis}}, \bibinfo {author} {\bibfnamefont {E.~D.}\ \bibnamefont {Kovetz}},
  \bibinfo {author} {\bibfnamefont {Y.}~\bibnamefont {Ali-Haïmoud}}, \bibinfo
  {author} {\bibfnamefont {S.}~\bibnamefont {Bird}}, \bibinfo {author}
  {\bibfnamefont {M.}~\bibnamefont {Kamionkowski}}, \bibinfo {author}
  {\bibfnamefont {J.~B.}\ \bibnamefont {Muñoz}}, \ and\ \bibinfo {author}
  {\bibfnamefont {A.}~\bibnamefont {Raccanelli}},\ }\href {\doibase
  10.1103/PhysRevD.94.084013} {\bibfield  {journal} {\bibinfo  {journal} {Phys.
  Rev.}\ }\textbf {\bibinfo {volume} {D94}},\ \bibinfo {pages} {084013}
  (\bibinfo {year} {2016})},\ \Eprint {http://arxiv.org/abs/1606.07437}
  {arXiv:1606.07437 [astro-ph.HE]} \BibitemShut {NoStop}%
%%CITATION = ARXIV:1606.07437;%%
\end{thebibliography}%

\end{document}